\documentclass[11pt]{article}
\usepackage[utf8]{inputenc}
\usepackage[T2A]{fontenc}
\usepackage{mathtext}
\usepackage{amssymb}
\usepackage{graphics}
\usepackage[dvips]{graphicx}
\usepackage{color}
\usepackage{amsbsy}
\usepackage{citehack}
\usepackage{longtable}
\usepackage[left=2cm,right=2cm,top=2cm,bottom=2cm,bindingoffset=0cm]{geometry}
\usepackage{amsmath}
\usepackage{ccaption}
\usepackage{epstopdf}
\usepackage{soul}
\usepackage{verbatim}
\usepackage{dsfont}

\usepackage{authblk}
\usepackage{stmaryrd}
\usepackage{appendix}
\usepackage{hyperref}
\numberwithin{equation}{section}

\title{Supersonic kinks and solitons in active solids}
\author{N.Gorbushin}
\author{L. Truskinovsky}
\affil{ \it Laboratoire de Physique et M\'ecanique des Milieux H\'et\'erog\'enes (PMMH UMR 7636) CNRS, ESPCI Paris, PSL Research University, 10 rue Vauquelin, 75005 Paris, France}

\date{}

\begin{document}
\maketitle
\begin{abstract}
To show that steadily propagating nonlinear waves in active matter can be driven internally,  we develop a prototypical model of a topological kink moving with a constant supersonic speed.  We use a  model of a bi-stable mass-spring (FPU)  chain   capable of  generating active stress. In contrast to subsonic kinks in   passive bi-stable  chains, that are necessarily dissipative, the obtained supersonic solutions are purely anti-dissipative. Our numerical experiments point  towards  stability of  the obtained  kink-type solutions and the possibility of  propagating kink-anti-kink   bundles reminiscent of solitons.  We show that  even the simplest quasi-continuum approximation of the  discrete model captures the  most important features of the predicted active   phenomena.
\end{abstract}

\section{Introduction}

Recently, considerable efforts  have been focused on the modeling of active matter.  We use this general term to describe a collection of interacting active agents,  each one driven by its own internally fuelled mechanism~\cite{marchetti2013hydrodynamics,de2015introduction,fodor2018statistical,berthier2019glassy}.
An important  question concerns nucleation and propagation of transition fronts in such systems  which can be modeled as \emph{kinks} separating passive and active phases.

 It was found that in various systems of  self-propelled active objects, ranging from   microorganisms to swarming robots, the transition from random to coherent motion is indeed accompanied by the formation of sharp  transition zones. They were observed to be moving with particular velocities whose  selection principle still remains an open problem~\cite{solon2015pattern,ngamsaad2018propagating}. Most of the related theoretical work was done using various coarse grained versions of the discrete Vicsek model where active agents are postulated to move with particular speeds. 
 \maketitle
 
In this paper we study kink-type solutions,  imitating transition fronts,  in a different model of active matter which, instead of supporting active velocities,  is able to generate  locally active stresses. The known continuum models of   active media with internally generated active stresses range from fluids ~\cite{prost2015active,julicher2018hydrodynamic} to solids~\cite{hawkins2014stress,maitra2018oriented,moshe2018geometric,scheibner2019odd}.
Behind these models is the idea of chemo-mechanical coupling~\cite{finlayson1969convective} with active stress emerging constitutively from the cross term linking a scalar chemical reaction with a tensorial mechanical action. For instance, in active gel theory the coupling is accomplished through an additional  liquid-crystal-type  vector field describing local polarization. 
 
\begin{figure}[h!]
\center{\includegraphics[scale=1.2]{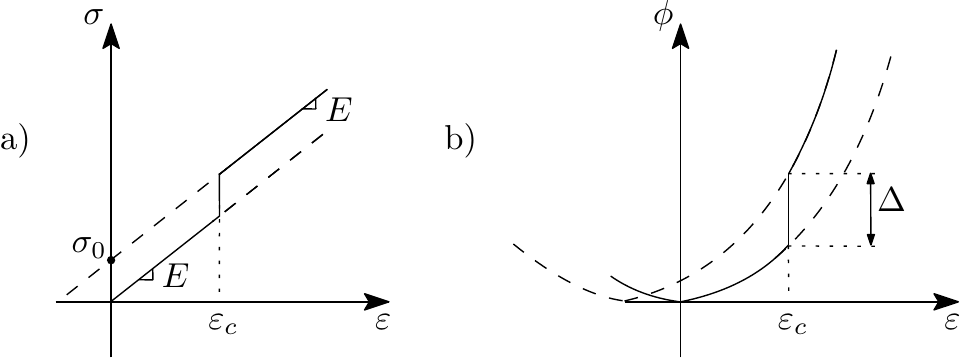}}
\caption[ ]  { Constitutive behavior of a continuum element which can undergo a transition from passive $\varepsilon<\varepsilon_c$ to active $\varepsilon>\varepsilon_c$ regime: (a)   stress-strain relation $\sigma=\sigma(\varepsilon)$ , (b)    free  energy  $\phi=\phi(\varepsilon)$.}
\label{fig:Constitutive}
\end{figure}

To achieve anlytical transparency  we   neglect the chemical fuelling side of the problem and  disregard polarization,  assuming that the prescribed active stress is hydrostatic. If we limit our description to 1D but, in view of the anomalous softness of the associated  phases, keep the inertial terms, we can write the ensuing   continuum problem for the displacement field $u(x,t)$ in the form: 
\begin{equation}
\label{eq:ShockProblem2}
\rho\frac{\partial^2 u}{\partial t^2}= \frac{\partial \sigma}{\partial x}, 
\end{equation}
where we denoted by $\rho$ the constant reference mass density. The activity is hidden in the constitutive relation for the stress $\sigma$ which we assume to be elastic and represented by two branches: passive, $$\sigma =
E\varepsilon,$$ where $\varepsilon(x,t)=\partial u(x,t)/\partial x$ is the strain and $E$ is the elastic modulus, and active, $$\sigma = 
E\varepsilon+\sigma_0,$$ where $\sigma_0>0$ is the active stress which is fixed. The corresponding branches of the energy density  are: $\phi=
(E/2)\varepsilon^2$  (passive) and  $\phi(\varepsilon)=
(E/2)\varepsilon^2+\sigma_0(\varepsilon-\varepsilon_c)+\Delta$ (active). Here, we introduced the critical strain 
$\varepsilon_c$ where the transition from passive to active branch   takes place;  such transition requires   energy expenditure (see Fig.~\ref{fig:Constitutive}):
$$\Delta=\sigma_0\varepsilon_c+\frac{\sigma_0^2}{2E},$$
which is assumed to be supplied actively at the microscopic level.
\begin{figure}[h!]
\center{\includegraphics[scale=1]{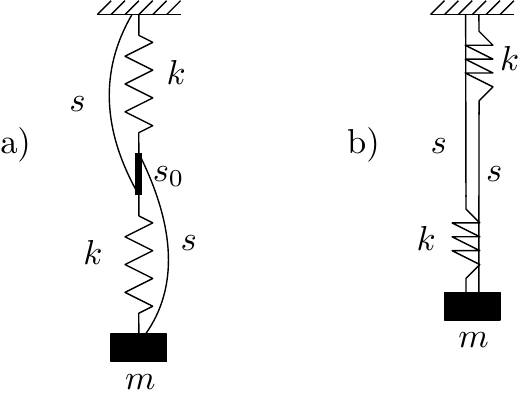}}
\caption[ ]{The mechanical model illustrating  the Braess paradox  \cite{cohen1991paradoxical}. When the link $s_0$ is cut the structure transform   series to parallel arrangement of springs and the mass is lifted.}
\label{fig:Paradox}
\end{figure}

To show that the resulting  material model  is microscopically  meaningful, consider a mechanical structure proposed in~\cite{cohen1991paradoxical} as an application of the Braess paradox from the game theory~\cite{braess1968paradoxon}. The system includes two elastic springs with elasticity $k$  and initial length $a$, two supporting inextensible elements  with the length $s$, a linking inextensible element with the length  $s_0$ and a mass $m$, see  Fig.~\ref{fig:Paradox}. When the string $s_0$ is cut the mass is lifted by the amount  $s_0-s+a+3mg/(2k)$,  where $g$ is the acceleration of gravity (the spring connection transforms  from series to parallel). The model depicted in Fig. \ref{fig:Constitutive} can be viewed as a stylized version of this mechanical system with the series  arrangement describing the passive branch of the constitutive relation while the  parallel arrangement corresponding to the active branch. The latter  'generates' extra stress which ultimately lifts the load. 

Note that the transition to the active branch requires energy expenditure because the implied  cutting  ultimately involves bond breaking. The corresponding energy "cost"    is characterized in our model by the parameter $\Delta$ and   the required  work can be viewed as produced by active forces implicitly present in the system. We note that the idea of~\cite{cohen1991paradoxical}  was recently used in~\cite{nicolaou2012mechanical} to build a passive metamaterial with negative compressibility. The interpretation of such material as active implies the presence of an active agent inserting the energy into the system each time the passive element reaches the critical strain and extracting it when the active material transforms back into its passive state; for a  passive model where microscopically stored energy is used to push a propagating front see~\cite{ayzenberg2014brittle}.

The focus of this paper is  on  velocity  \emph{selection} for the  switching   waves  transforming the passive (analog of Vicsek's disordered) state into the active (analog of of Vicsek's ordered) state.  Such waves are similar to the topological kinks in passive bi-stable materials except that they are not  driven   externally, by the applied stress biasing the double well potential, but  internally,  as a result of the energy consumption   from the macroscopically invisible out of equilibrium chemical reservoir.  

We first use the classical continuum theory to show that such kinks are necessarily   supersonic and that their velocity remains undefined unless the microscopic problem is  solved first.  We then  solve analytically the corresponding discrete problem  for a bi-stable FPU (Fermi-Pasta-Ulam) chain~\cite{gallavotti2007fermi} 
and  obtain   the required  kinetic relation explicitly. It turns out to be universal but trivial.  

More precisely, in contrast to subsonic kinks in a passive bi-stable FPU system that are necessarily dissipative, the obtained supersonic kinks are dissipation free. Our numerical experiments point  towards stability of such waves. We also show  the possibility of  propagating kink-anti-kink   bundles reminiscent of solitons and show that such active solitary waves indeed collide almost elastically.  We also check  that  the  simplest quasi-continuum approximation of the  discrete model captures the most important features of the active kink propagation  phenomenon both qualitatively and quantitatively.
 
Our main technical tool is the Fourier transform which can be used in this nonlinear problem due to  the piece-wise linear, equal moduli  approximation of the bi-stable constitutive relation. Similar approaches have been previously used  for the description of the transition waves in passive FPU type systems \cite{truskinovsky2005kinetics,slepyan2005transition,
truskinovsky2006quasicontinuum}. To treat the case of different moduli, we could have used only  slightly  more complex approach   based on the Wiener-Hopf  method   \cite{slepyan1988impact, slepyan2012models}, however,  we have checked that such augmentation  of the model does not change any of our main results.

The paper is organized as follows. In §~\ref{sec:Continuous model}, we introduce a continuum model and show  that it  produces non-unique solutions and does not select a particular  kink velocity. In §~\ref{sec:Discrete model}, we develop the   discrete model and solve its explicitly. We then show numerically the stability of the obtained solutions. The simplest  quasi-continuum approximation of the discrete model is developed  in §~\ref{sec:Quasi-continuum model}. Finally, in §~\ref{sec:Active solitary waves}, we present numerical evidence that kinks and anti-kinks can bundle together to form soliton like localized solutions. We present our  conclusions in §~\ref{sec:Conclusions}. The solutions describing analytically tractable    kinks  in the linear discrete problem  are presented in appendix~\ref{sec:Sonic wave}.

\section{Continuum model}
\label{sec:Continuous model}

Staying within the continuum setting,  we can  rewrite our second order  dynamic equation \eqref{eq:ShockProblem2} as the first order system:
\begin{equation}
\frac{\partial \varepsilon}{\partial t}=\frac{\partial v}{\partial x},\quad \rho \frac{\partial v}{\partial t}=\frac{\partial \sigma(\varepsilon)}{\partial x},
\label{eq:ShockProblem}
\end{equation}
where  $v(x,t)=\partial u(x,t)/\partial t$ is the velocity field and 
\begin{equation}
\sigma(\varepsilon)=
\begin{cases}
E\varepsilon,\quad \varepsilon<\varepsilon_c,\\
E\varepsilon+\sigma_0,\quad \varepsilon>\varepsilon_c 
\end{cases} 
\label{eq:Stress_Energy}
\end{equation} 
is the stress-strain relation.  We are interested in the behavior of sharp discontinuities mimicking (diffuse) transition fronts. On the corresponding fronts the equations of elastodynamics \eqref{eq:ShockProblem} have to be supplemented by the  Rankine-Hugoniot jump conditions:
\begin{equation}
\llbracket v \rrbracket + V\llbracket \varepsilon \rrbracket=0,\quad \rho V\llbracket v \rrbracket+\llbracket \sigma(\varepsilon)\rrbracket=0,
\label{eq:R-H_conditions}
\end{equation}
where $\llbracket f \rrbracket\equiv f_+ - f_-$ is the difference between the limiting values $f_+$ and $f_-$ of a function $f(x)$  from the  the right and left of the discontinuity.  

While the phenomena in the bulk described by the system \eqref{eq:ShockProblem} are  necessarily non-dissipative,   discontinuities may serve as potential sources of dissipation with classical shock  waves as well known examples. To address this issue we need to write the  integral energy balance equation 
\begin{equation}
\sigma v \bigr\rvert^{+\infty}_{-\infty} +V \Delta - \frac{d}{dt}\int_{-\infty}^{\infty}\left[\frac{v^2}{2}+\phi(\varepsilon)\right]=GV, 
\label{eq:EntropyCondition}
\end{equation}
where $ G(V)=\llbracket \phi(\varepsilon) \rrbracket-\{\sigma(\varepsilon)\}\llbracket \varepsilon \rrbracket$ is the driving/configurational  force  acting on the discontinuity, $\{f\}=(f_++f_-)/2$, and  we assumed for determinacy that the body is infinite~\cite{truskinovskii1987dynamics}. Note the unusual term $V \Delta $ which stands for the work performed  by the  \emph{active agency} in transforming the system from   passive  to   active state. Since our model is purely mechanical, the analog of the second law of thermodynamics, implying that the energy flux from macro-scale to micro-scale is irreversible, amounts to the requirement that 
\begin{equation}
GV\geq 0,
\label{eq:EntropyCondition1}
\end{equation}
which can serve as an additional (entropic) jump condition. 
 
\begin{figure}[h!]
\center{\includegraphics[scale=1. ]{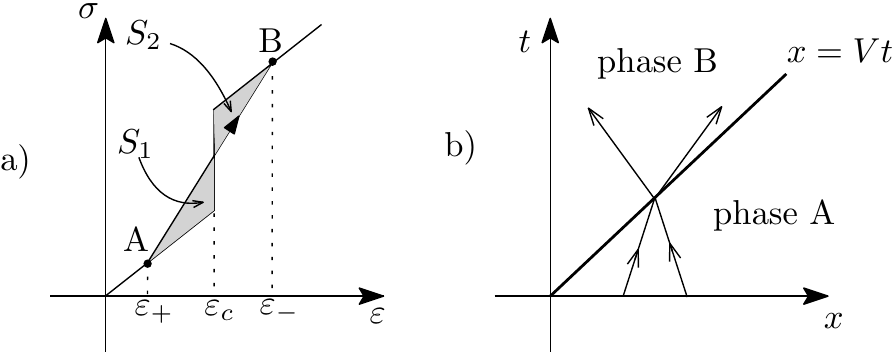}}
\caption[ ]{(a) The stress-strain relation crossed by  the Rayleigh line $\sigma-\sigma_+=\rho V^2(\varepsilon-\varepsilon_+)$
alows one to define the driving force as the area difference $G=S_2-S_1$, (b) the $(x,t)$ diagram showing the moving front interacting with  two incoming and two outgoing characteristics.}
\label{fig:Characteristics}
\end{figure}
 
To check whether the obtained set of jump conditions  \eqref{eq:R-H_conditions},\eqref{eq:EntropyCondition1}  is sufficient, we need to specify the type of the discontinuity, see  Fig.~\ref{fig:Characteristics}a).  Consider the  transition from a generic  passive state  A,  with  $\varepsilon=\varepsilon_+$, to the active state B, with  $\varepsilon=\varepsilon_-$; the corresponding  transformation front (strain discontinuity) traveling at a constant speed $V>0$ transforms the state  with the constant velocity $v=v_+$ to the state  with the constant velocity $v=v_-$.  

Note that the  front AB is necessarily \emph{supersonic} with respect to the  state ahead of it, $V>c_+$,   and is   also \emph{supersonic} with respect to   state  behind it,   $V>c_-$, where    we introduced the characteristic  speeds   $c_{\pm}=\sqrt{\sigma'(\varepsilon_{\pm})/\rho}$. Therefore, if we consider a  point $(x,t)$ which coincides with the current position of the front, two characteristics moving with velocities $\pm c_+$ will be coming to the front from the side of state A and two will be leaving to the side of state B, see  Fig.~\ref{fig:Characteristics}b). Note also that for the classical phase boundaries, which are {\it subsonic} with respect to the state ahead and {\it subsonic} with respect to the state behind, there will be also two coming and two leaving characteristics, even though one of the coming characteristics will be   arriving from behind, see~\cite{truskinovsky1993kinks}. In both cases, however, with two arriving characteristics, two Rankine-Hugoniot jump conditions \eqref{eq:R-H_conditions},  and the condition \eqref{eq:EntropyCondition} being just an inequality, we do not have enough data to specify the 5 unknown parameters: states ahead and behind, $(\varepsilon_{\pm},v_{\pm})$, plus the velocity of the discontinuity $V$.

To find the missing condition, which ultimately selects the velocity of the front,  we need to solve the corresponding microscopic problem, which should allow one not only to confirm  the non-negativity of the product $ GV$ but also to specify the \emph{kinetic relation} $G=f(V)$. 

\section{Discrete model}
\label{sec:Discrete model}
Consider now the simplest  mass-spring chain  imitating the behavior of the continuum system \eqref{eq:ShockProblem2}. To this end we need to assume that the  non-linear springs obey the force-elongation (stress-strain) relation $\sigma(\varepsilon)$ given by \eqref{eq:Stress_Energy}, see Fig.~\ref{fig:Constitutive}a).  Suppose also that the equilibrium position of the masses are $x_n=na$ where $a$ is the equilibrium spring length and $n$ is an integer.  

The dynamics of such FPU  chain is governed by the   equations: 
\begin{equation}
\rho a \frac{d^2 u_n(t)}{dt^2}=\sigma\left(\frac{u_{n+1}-u_{n}}{a}\right)-\sigma\left(\frac{u_{n}-u_{n-1}}{a}\right),
\label{eq:EquationsOfMotion_displ}
\end{equation}
where $\rho$ is the mass density and $u_n(t)$ is the displacement of the $n$-th mass.   Introducing the discrete strain $\varepsilon_n(t)=(u_{n+1}(t)-u_n(t))/a$, we can rewrite  this system in the form:

\begin{equation}
\begin{gathered}
\rho a^2\frac{d^2 \varepsilon_n(t)}{dt^2}=\sigma(\varepsilon_{n+1})+\sigma(\varepsilon_{n-1})-2\sigma(\varepsilon_{n}).\\
\end{gathered}
\label{eq:EquationsOfMotion}
\end{equation}

We are interested in the   traveling wave (TW) solutions of the equations \eqref{eq:EquationsOfMotion}. We therefore assume that:
\begin{equation}
\varepsilon_n(t)=\varepsilon(\eta),\quad \varepsilon_{n\pm 1}(t)=\varepsilon(\eta\pm a),\quad \eta=na-Vt.
\label{eq:Strain_a}
\end{equation}
The point $\eta=0$ will be associated with the position of   the moving front. 

If we   non-dimensionalize the TW problem by introducing new variables: 
$$
\hat{V}= V/c, \quad \hat{\eta}= \eta/a, \quad \hat{\sigma}= \sigma/E,\quad \hat{\sigma}_0= \sigma_0/E,\quad \hat{\phi}= \phi/E 
$$
and then drop  the hats for simplicity, we obtain 
a single equation of motion in  the  form  
\begin{equation}
V^2\frac{d^2\varepsilon}{d\eta^2}=\sigma(\eta+1)+\sigma(\eta-1)-2\sigma(\eta).
\label{eq:equations_steady}
\end{equation}
where $$
\sigma(\eta)=\varepsilon(\eta)+\sigma_0H(-\eta),
$$
and  $H(\eta)$ is the Heaviside   function. 
For consistency with our stress-strain relations we need to  supplement the equation \eqref{eq:equations_steady} by the switching condition:
\begin{equation}
\varepsilon(0)=\varepsilon_c.
\label{eq:Condition}
\end{equation}
The transitional wave (kink) must also  satisfy the following boundary conditions: 
\begin{equation}
\varepsilon(\eta)=
\begin{cases}
\varepsilon_+,\quad \eta\to\infty,\\
\varepsilon_-,\quad \eta\to-\infty
\end{cases}
\label{eq:BCs}
\end{equation}
The conditions  \eqref{eq:BCs} should be understood in the sense of averages in view of the possibility of the macroscopically invisible lattice wave oscillations carrying the energy away from the discontinuity \cite{slepyan2012models}. 

It is natural to look for the solution  of the problem in the form $$\varepsilon(\eta)=\mathcal{E}(\eta)+\varepsilon_+$$ and   instead of  $\mathcal{E}(\eta)$ it will be more convenient to consider its Fourier transform 
$$
\check{\mathcal{E}} (p)=\int_{-\infty}^{\infty}\mathcal{E}(\eta)e^{ip\eta}\,d\eta.
$$
In the Fourier space the equation \eqref{eq:equations_steady}  reduces to:
\begin{equation}
L(p)\check{\mathcal{E}}(p)=-\sigma_0\frac{\omega^2(p)}{(0+ip)}, 
\label{eq:Fourier}
\end{equation}
where 
\begin{equation}
L(p)=\omega^2(p)-(pV)^2
\label{eq:Fourier1}
\end{equation}
is the Fourier image of the linear operator describing dynamics in each of the phases and 
\begin{equation}
\omega^2(p)=4\sin^2\left(\frac{p}{2}\right)
\label{eq:Fourier2}
\end{equation}
is the dispersion relation in the linearized model. The denominator $(0+ip)$ in \eqref{eq:Fourier} comes from the Fourier transform of the Heaviside function and should be understood as $\lim_{\alpha\to0+}(\alpha+ip)$.  In what follows  we write directly $ip$ instead of $0+ip$ keeping in mind that the singularity at $p=0$ should be understood in the sense of the above limit.

The solution of  \eqref{eq:Fourier} is  straightforward and by inverting the Fourier transform we obtain  
\begin{equation}
\varepsilon(\eta)=\varepsilon_+-\frac{\sigma_0}{2\pi i}\int_{-\infty}^{\infty}\frac{\omega^2(p)}{L(p)}\frac{e^{-ip\eta}}{p}\,dp.
\label{eq:Solution}
\end{equation}
\begin{figure}[h!]
\center{\includegraphics[scale=0.35]{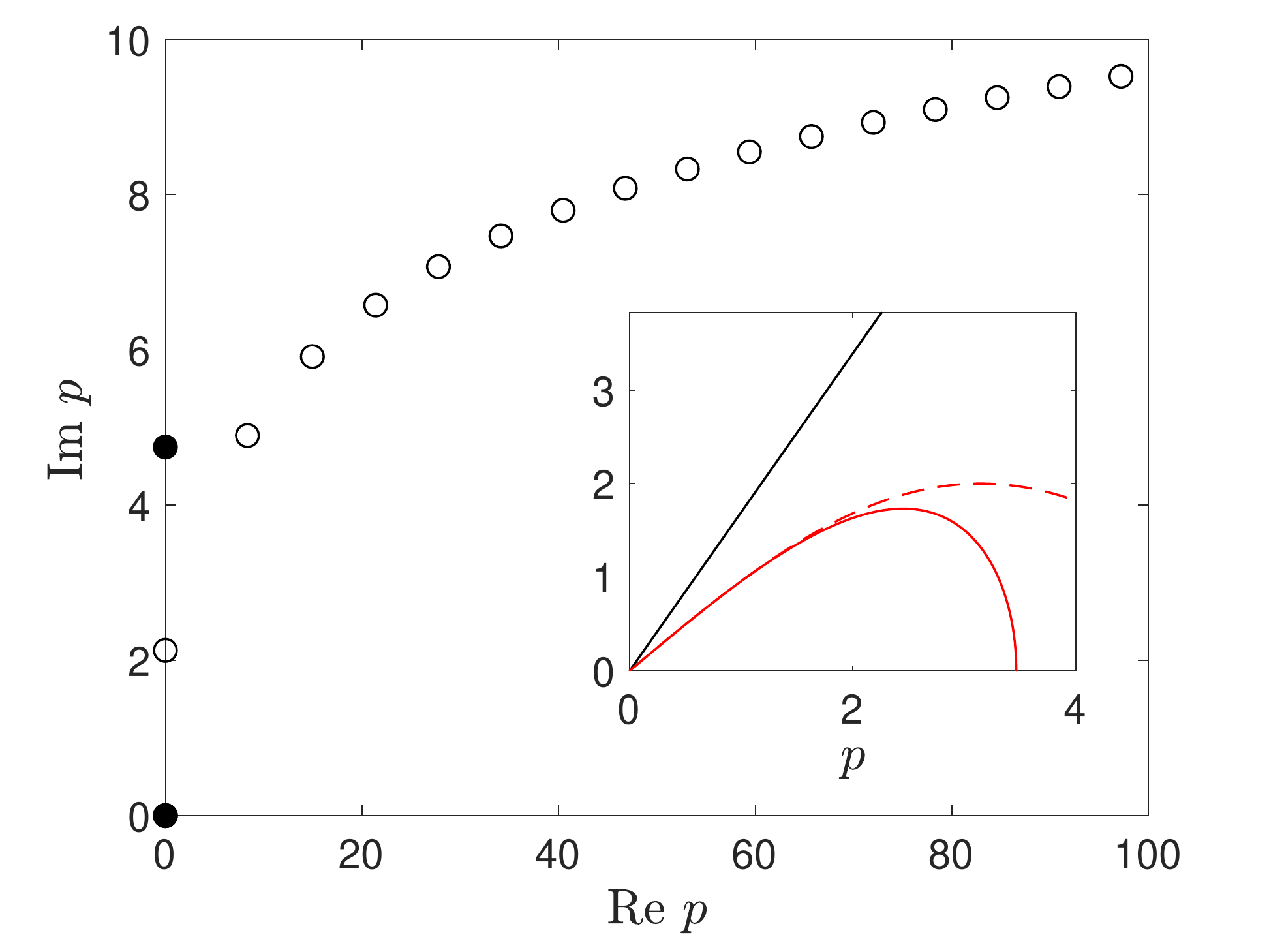}}
\caption[ ]{Zeros of the function $L(p)$, see  \eqref{eq:Fourier1},  inside the first quadrant for the discrete (hollow circles) and quasi-continuum (filled circles) models. The inset shows the dispersion relation $\omega(p)$, for the discrete (red dashed line), see \eqref{eq:Fourier2},  and quasi-continuum (red solid line), see \eqref{fig:quasicont},  models in relation to the line  $\omega=Vp$ (black solid line).}
\label{fig:Dispersion_relation}
\end{figure}
To evaluate   the integral in \eqref{eq:Solution} we can use the residue theorem. Due to symmetry,  purely imaginary roots of  the function $L(p)$ appear in pairs while  complex roots appear in quadruplets~\cite{truskinovsky2005kinetics}.  Since $V>1$, the curves $\omega(p)$ and $Vp$ have no intersections except at $p=0$ and the real roots are absent (Fig.~\ref{fig:Dispersion_relation}). 

To specify the function $\varepsilon(\eta)$ we need to introduce the following sets 
\begin{equation}
Z^\pm=\left\{p\,:\,L(p)=0,\,\pm\text{Im } p>0\right\}.
\end{equation}
We can then  write the explicit solution of the problem in the form 
\begin{equation}
\begin{gathered}
\varepsilon(\eta)=
\begin{cases}
\varepsilon_++\sum\limits_{p_j\in Z^-}\dfrac{\sigma_0\omega^2(p_j)}{p_jL'(p_j)}e^{-ip_j\eta},\quad \eta>0,\\
\varepsilon_++\dfrac{\sigma_0}{V^2-1}-\sum\limits_{p_j\in Z^+}\dfrac{\sigma_0\omega^2(p_j)}{p_jL'(p_j)}e^{-ip_j\eta},\quad \eta<0.
\end{cases}
\end{gathered}
\label{eq:Solution_series}
\end{equation}
where we used  the fact that  the point $p=0$ should be  passed by the integration contour from below, see \cite{truskinovsky2005kinetics} for more details. 
\begin{figure}[h!]
\begin{center}
a)\hspace{-0.5cm}\begin{minipage}[h]{0.35\linewidth}
\center{\includegraphics[scale=0.25]{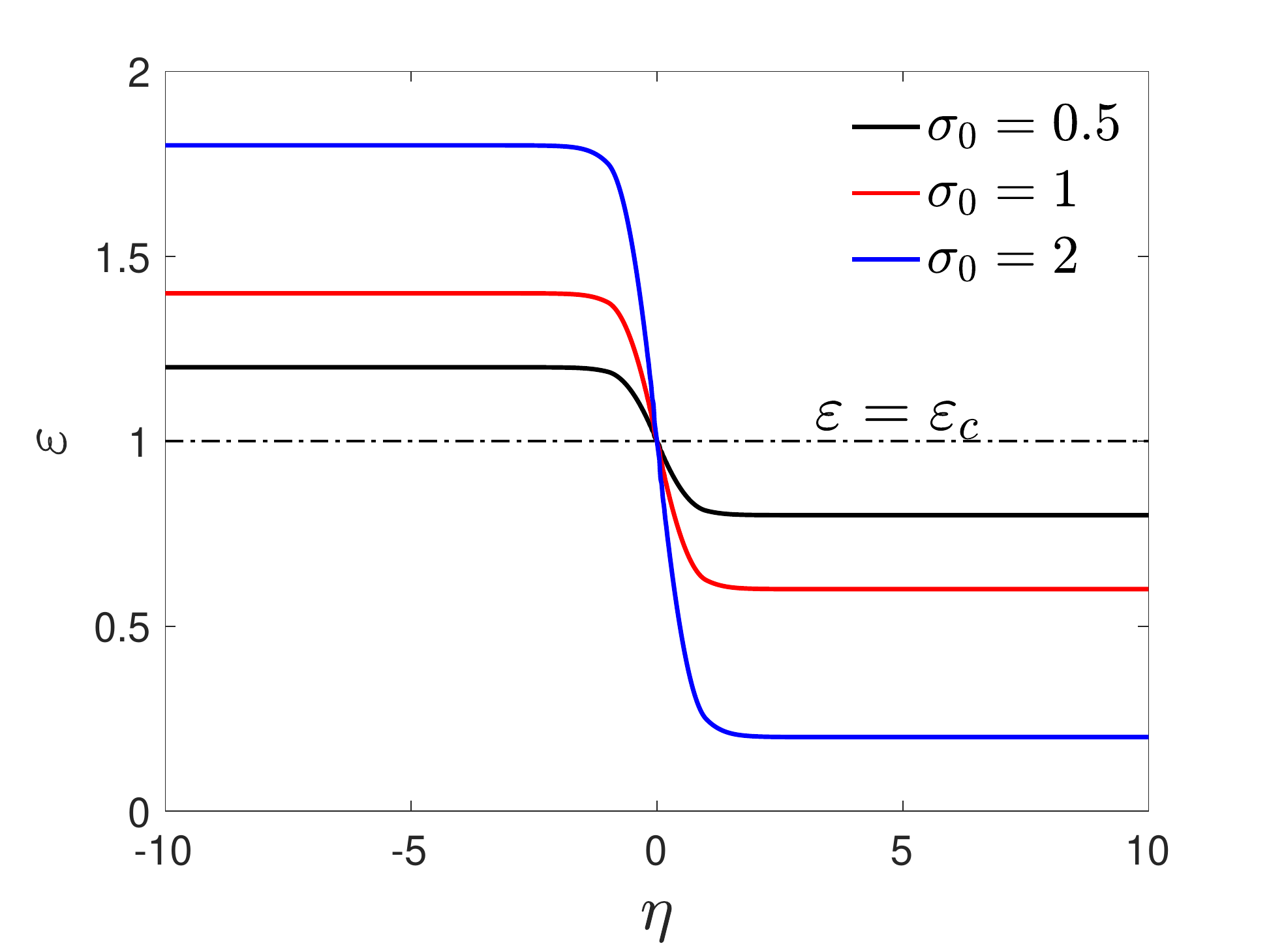}}
\end{minipage}
b)\hspace{-0.5cm}\begin{minipage}[h]{0.35\linewidth}
\center{\includegraphics[scale=0.25]{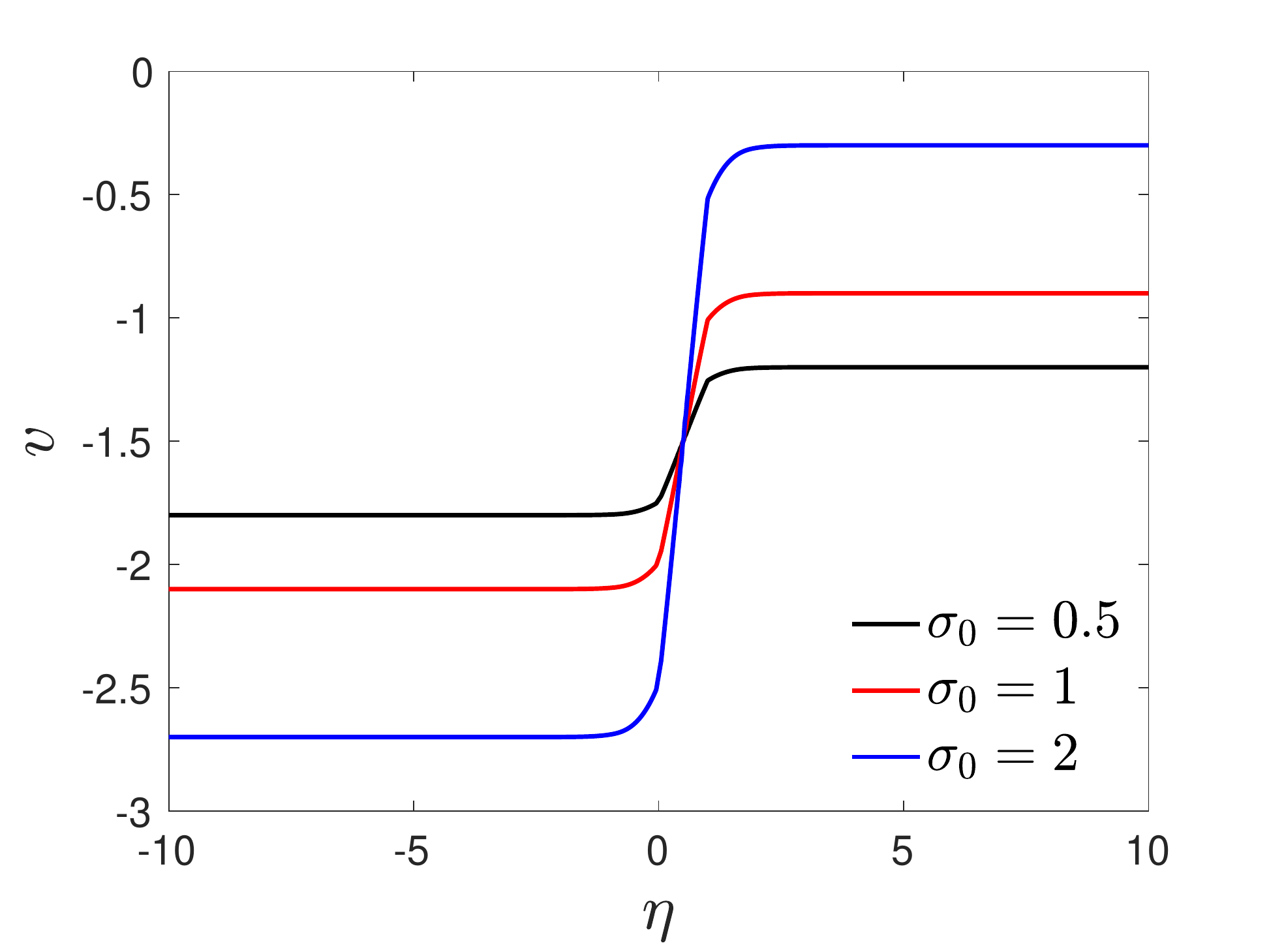}}
\end{minipage}
\end{center}
\caption[ ]{Strain (a) and velocity (b)  profiles at $V=1.5$ and different $\sigma_0$.}
\label{fig:Solution}
\end{figure}

\begin{figure}[h!]
\center{\includegraphics[scale=0.25]{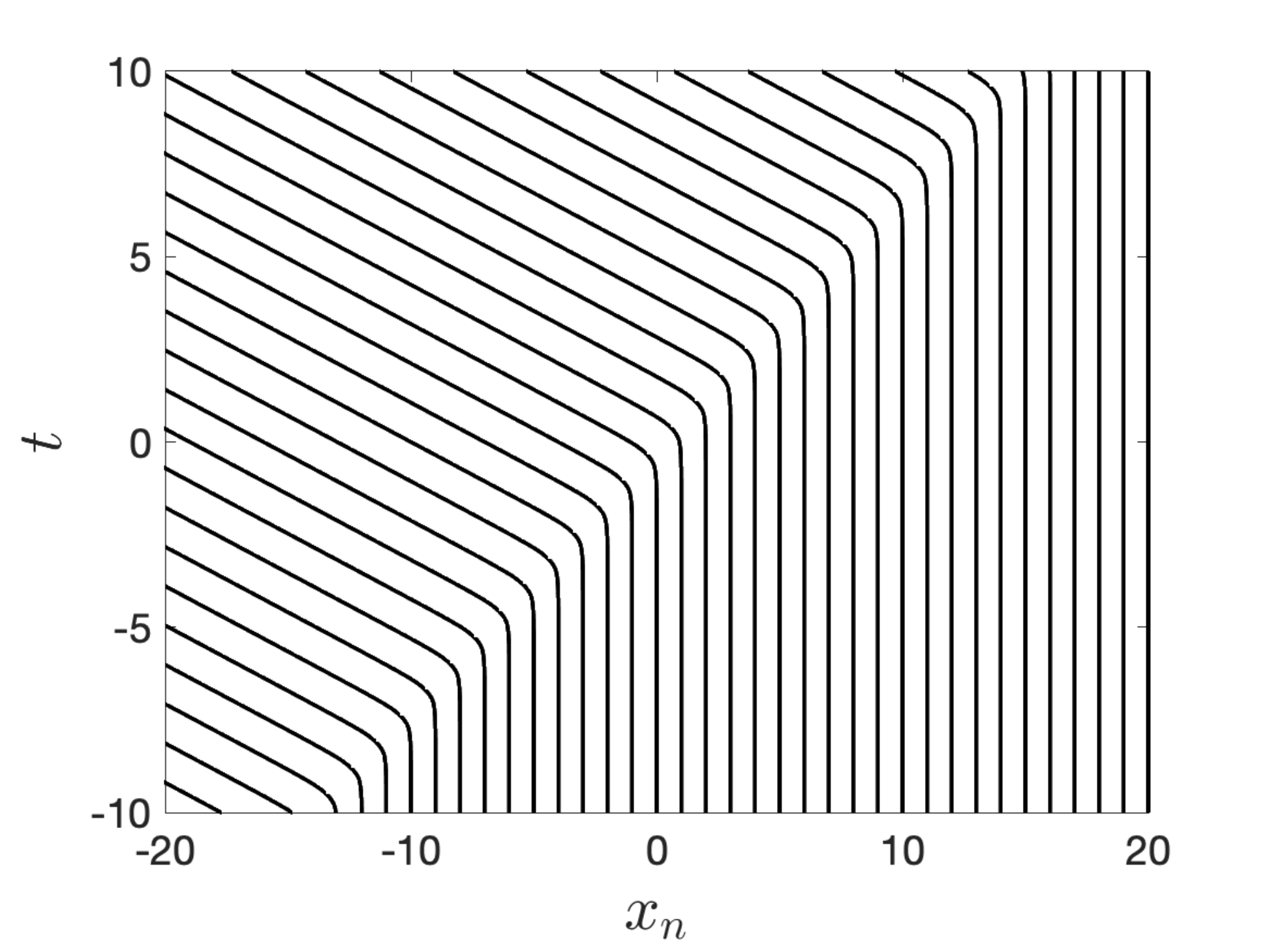}}
\caption[ ]{Trajectories of mass points  $x_n(t)=n+u_n(t)$ for the kink moving with the velocity $V=\sqrt{2}$;   $\sigma_0=2$.}
\label{fig:Trajectories}
\end{figure}

The typical strain profiles are shown in Fig.~\ref{fig:Solution}a); these   solutions are   admissible in the sense that the   critical strain is passed only once.  To find the  velocity distribution $v(\eta)$ we need to solve the equation
$
v(\eta)=-V d u(\eta)/d \eta,
$
which in the Fourier space reads  
\begin{equation}
\check{\mathcal{V}}(p)=-Ve^{ip/2}\frac{p/2}{\sin{(p/2)}}\left[2\pi\varepsilon_+\delta(p)+\check{\mathcal{E}}(p)\right]. 
\end{equation}
In the real space we obtain (see Fig. \ref{fig:Solution}b))
\begin{equation}
v(\eta)=
\begin{cases}
-V\varepsilon_+-\frac{V}{2}\sum\limits_{p_j\in Z^-}\dfrac{\sigma_0\omega^2(p_j)}{\sin{(p_j/2)}L'(p_j)}e^{-ip_j(\eta-1/2)},\quad \eta>1/2,\\
-V\varepsilon_- +\frac{V}{2}\sum\limits_{p_j\in Z^+}\dfrac{\sigma_0\omega^2(p_j)}{\sin{(p_j/2)}L'(p_j)}e^{-ip_j(\eta-1/2)},\quad \eta<1/2.
\end{cases}
\label{eq:Solution_velocity}
\end{equation}
The typical velocity profiles are shown in Fig.~\ref{fig:Solution}b).  Finally, to obtain the displacement field $u(\eta)$ we need to solve the equation $\varepsilon(\eta)=u(\eta+1)-u(\eta)$.  We can again use the Fourier transform to rewrite this equation in the form:
\begin{equation}
\check{U}(p)=\frac{i/2}{\sin{(p/2)}}\left(2\pi\varepsilon_+\delta(p)+\check{\mathcal{E}}(p)\right)
\end{equation}
By inverting the Fourier transform we obtain
\begin{equation}
u(\eta)=
\begin{cases}
\varepsilon_+\eta+\dfrac{i}{2}\sum\limits_{p_j\in Z^-}\dfrac{\sigma_0\omega^2(p_j)}{p_j\sin{(p_j/2)}L'(p_j)}e^{-ip_j(\eta-1/2)},\quad \eta>1/2,\\
\varepsilon_-\eta-\dfrac{\sigma_0}{2(V^2-1)} -\dfrac{i}{2}\sum\limits_{p_j\in Z^+}\dfrac{\sigma_0\omega^2(p_j)}{p_j\sin{(p_j/2)}L'(p_j)}e^{-ip_j(\eta-1/2)},\quad \eta<1/2.
\end{cases}
\label{eq:Solution_displ}
\end{equation}
where the linear term is due to a double pole at $p=0$;  an additive constant here  is set to 0. The resulting particle trajectories are illustrated in Fig.~\ref{fig:Trajectories}  for the case when $\varepsilon_+=0$ and $v_+=0$.

To compare the discrete solution (\ref{eq:Solution_series}, \ref{eq:Solution_velocity}, \ref{eq:Solution_displ}) with its piecewise constant continuum counterpart,  we first note  that by considering the limit $\eta\to\pm\infty$ in \eqref{eq:Solution_velocity}  we obtain 
$
v_{\pm}=-V\varepsilon_{\pm},
$
which  can be rewritten as the first Rankine-Hugoniot condition \eqref{eq:R-H_conditions}:
$
\llbracket v \rrbracket=-V\llbracket \varepsilon \rrbracket.
$
Note also that the conditions \eqref{eq:BCs} are fulfilled if 
\begin{equation}
\varepsilon_-=\varepsilon_++\frac{\sigma_0}{V^2-1}.
\label{eq:Epsilon_minus}
\end{equation}
which can be rewriten as the second Rankine-Hugoniot condition: 
$
V^2\llbracket \varepsilon \rrbracket-\llbracket \sigma(\varepsilon) \rrbracket=0.
$

Finally, we can use   \eqref{eq:Condition} to obtain the selection condition for the discontinuity velocity   $V$ (kinetic relation).   If we integrate $\omega^2(p)/(pL(p))$ over the   circle with the radius expanding to infinity and  apply  the residue theorem, we obtain
\begin{equation}
\sum\limits_{p_j\in Z}\frac{\omega^2(p_j)}{p_jL'(p_j)}=\frac{1}{V^2-1},
\label{eq:Sums}
\end{equation}
where the sum is taken over all zeros of $L(p)$. Hence, the continuity of $\varepsilon(\eta)$ at the point $\eta=0$ holds and the value $\varepsilon(0)=\varepsilon_c$ can be obtained by considering for instance the limit $\eta\to+0$ in \eqref{eq:Solution_series}. Moreover, due to the symmetry of the roots   we obtain
\begin{equation}
\sum\limits_{p_j\in Z^+}\frac{\omega^2(p_j)}{p_jL'(p_j)}=\sum\limits_{p_j\in Z^-}\frac{\omega^2(p_j)}{p_jL'(p_j)}
\end{equation}
which allows us to rewrite  the condition \eqref{eq:Condition} in the form
\begin{equation}
\varepsilon_+=\varepsilon_c-\frac{1}{2}\frac{\sigma_0}{V^2-1}.
\label{eq:Epsilon_pm}
\end{equation}
Equation \eqref{eq:Epsilon_pm} is the desired kinetic relation. It is easy to check that it is equivalent to the trivial condition 
\begin{equation}
G(V)= 0.
\label{eq:G_result}
\end{equation}
In other words, the magnitude of the driving force always takes the lowest possible value (equal to zero) and as a result the constructed solutions are    dissipation free.  Therefore,  all the energy provided by the active agency (which can be also interpreted as anti-dissipation) is consumed by the transformation itself. The absence of the radiative damping in the form of emitted lattice-scale waves  is the consequence of the supersonic nature of the transition which  effectively \emph{overtakes} any emitted acoustic wave. 

\begin{figure}[h!]
\center{\includegraphics[scale=0.3 ]{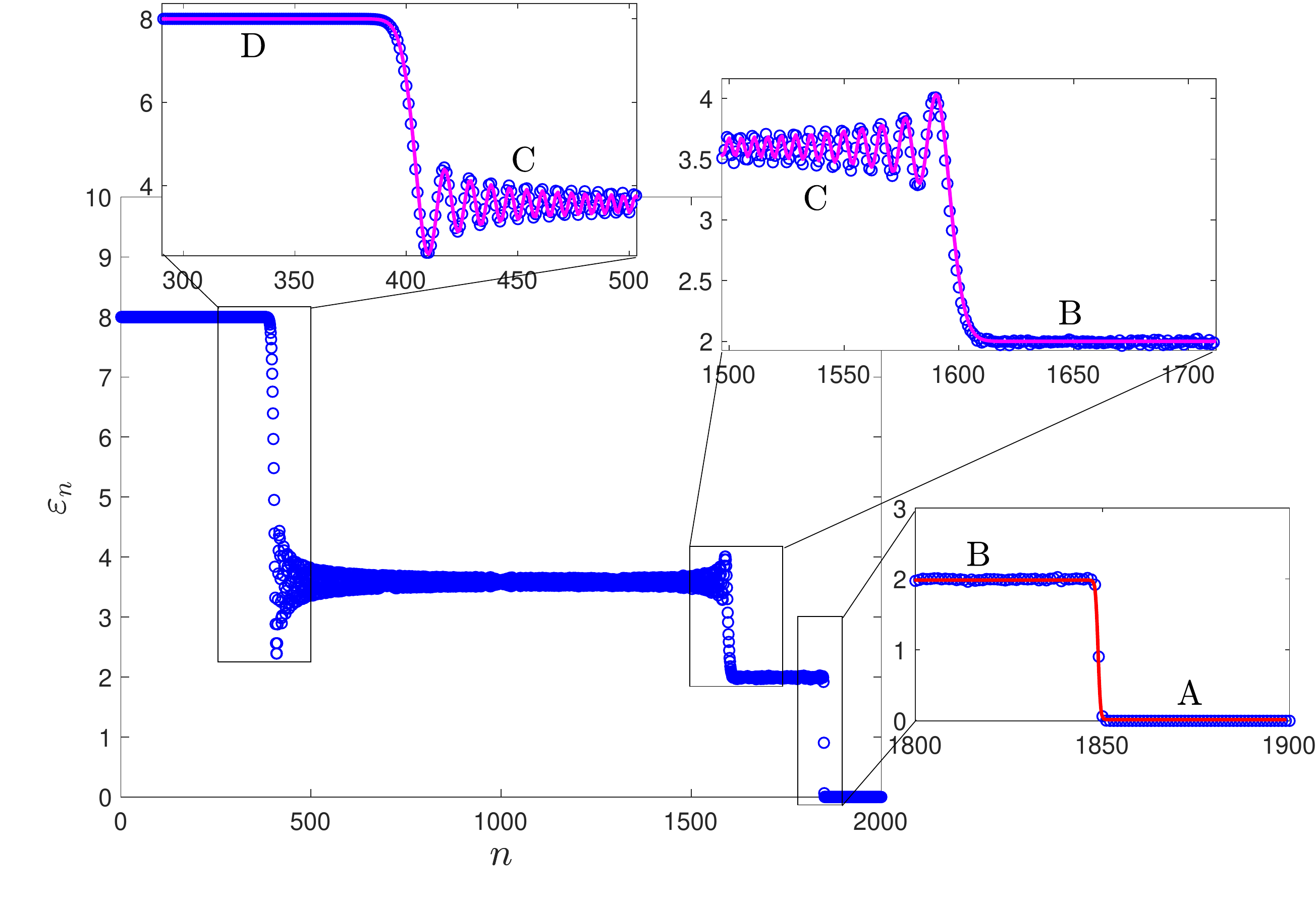}}
\caption[ ]{Results of numerical simulations with the Riemann initial conditions $(\varepsilon_l,\varepsilon_r)=(8,0)$ at $t=600$. The steady-state solutions in the right zoomed area is compared with the analytical solution \eqref{eq:Solution_series} (solid red line). The remaining inserts show expanding sonic waves matched with the theoretical solution \eqref{eq:SolutionGreen} (solid magenta lines). Different states A, B, C and D are indicated in the insets.}
\label{fig:Numerics_1}
\end{figure}

To check stability of the obtained solutions we performed numerical integration of the initial value problem for the system \eqref{eq:EquationsOfMotion} with $N=2000$ springs, $\varepsilon_c=1$ and  $\sigma_0=2$. In the first type of   tests we considered  the Riemann problem with initial conditions:
\begin{equation}
\left(\varepsilon_n(0),\frac{d\varepsilon_n}{dt}(0)\right)=\begin{cases}
(\varepsilon_l,0),\quad n<1000,\\
(\varepsilon_c,0),\quad n=1000,\\
(\varepsilon_r,0),\quad n>1000,
\end{cases}
\label{eq:RiemannData}
\end{equation}
where $\varepsilon_l$ and $\varepsilon_r$ are constants, while keeping  both ends of the chain   free. The results of a typical simulation of this type    are illustrated  in Fig.~\ref{fig:Numerics_1} where we show the time section  $t=600$ when all the fronts have sufficiently stabilized. We observe the  emergence of a steadily moving transition front AB  whose internal structure is in excellent agreement with the analytical solution \eqref{eq:Solution_velocity}, see the right inset in Fig.~\ref{fig:Numerics_1}. The steady-state traveling wave propagates  with the  velocity  $V=1.42$  which is  independent of the initial value $\varepsilon_l$ once we fixed $\varepsilon_r=0$. 
\begin{figure}[h!]
\center{\includegraphics[scale=1]{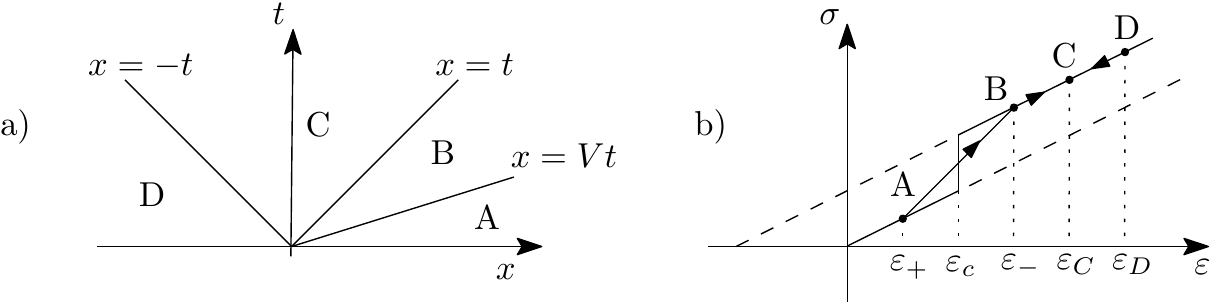}}
\caption[ ]{Breakdown of unstable piecewise constant initial state (Riemann problem): (a) Propagating  supersonic front  of nonlinear (passive-active) transformation  and two 'supporting' the sonic waves, (b) Global structure of the solution of the Riemann problem. The  analytical solutions \eqref{eq:Solution_series} (solid red line) and \eqref{eq:SolutionGreen} (solid magenta line) are  shown in insets for comparison.}
\label{fig:Disintegration}
\end{figure}

Note the presence of  two spreading sonic waves (BC and DC) which move in the opposite directions with the same speed $V=1$. Their internal structure is  reconstructed analytically in appendix~\ref{sec:Sonic wave} and our two other insets   show an excellent agreement between the theory and the numerical experiment. 

To interpret the global structure of the solution of this Riemann problem we show  in Fig.~\ref{fig:Disintegration}a) the spatial configuration of the all three propagating fronts.  The nonlinear transition AB and the linear transitions BC and CD are also illustrated  on the stress-strain curve in Fig.~\ref{fig:Disintegration}b) where  the average strains in the points  A and B are $\varepsilon_+$ and $\varepsilon_-$, respectively; within the active phase, we observe two strain plateaus  at  $\varepsilon_C$ and $\varepsilon_D=\varepsilon_l$. 
\begin{figure}[h!]
\center{\includegraphics[scale=0.35]{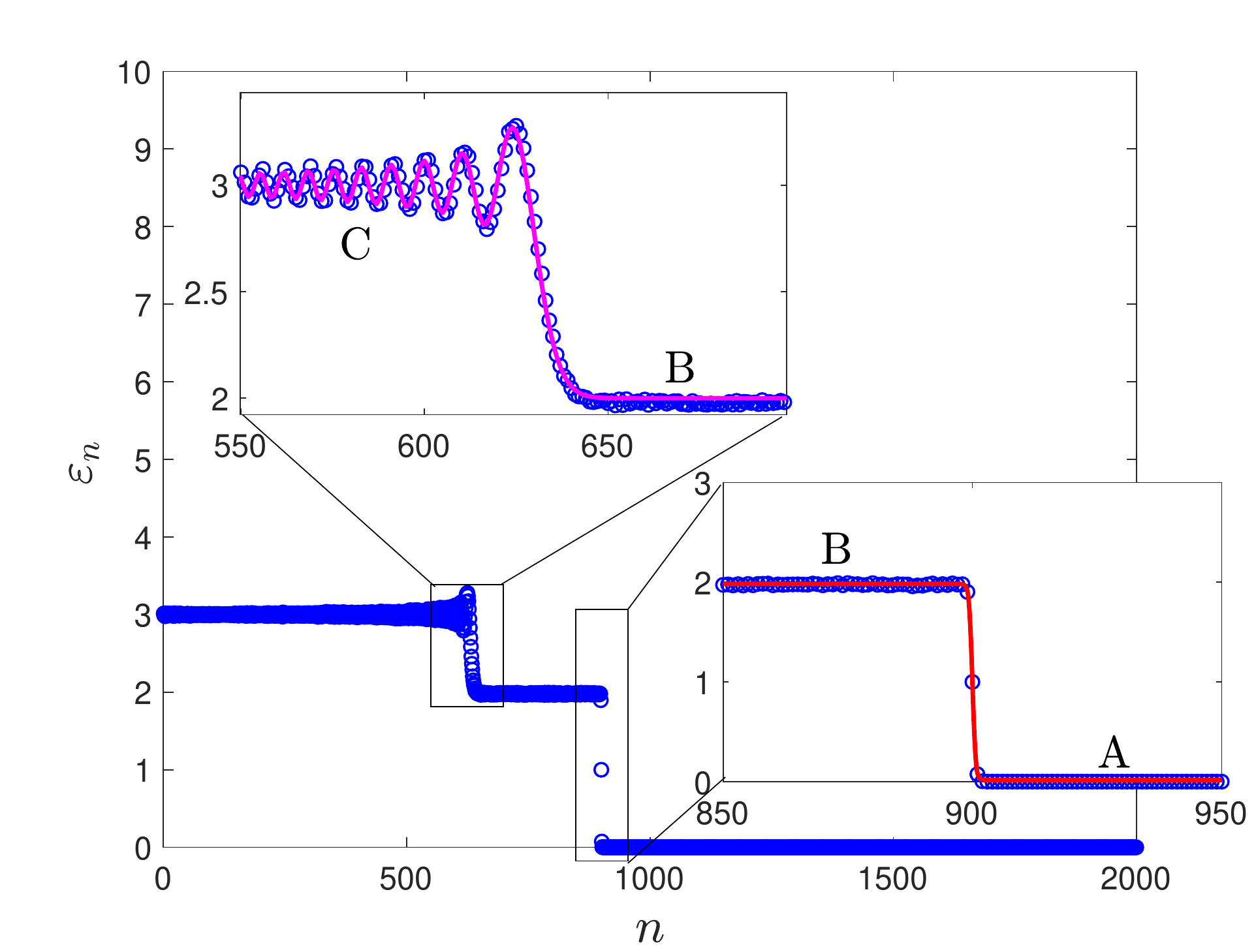}}
\caption[ ]{Snapshot of the numerical solution of the initial value problem with externally applied force $F=5$.  Here  $t=634$ when the first $900$ switching  transitions occurred. The  analytical solutions \eqref{eq:Solution_series} (solid red line) and \eqref{eq:SolutionGreen} (solid magenta line) are  shown in insets  for comparison.}
\label{fig:Numerics_2}
\end{figure}

We also  performed another type of numerical experiment where we solved equations \eqref{eq:EquationsOfMotion_displ}  with homogeneous initial conditions $u_n(0),\,du_n(0)/dt=0$ for any $n$,  while keeping the right end of the chain free and  applying a constant force $F$  at its left end: 
\begin{equation}
\left(u_n(0),\frac{du_n}{dt}(0)\right)=(0,0),\quad \frac{d^2u_1}{dt^2}=\sigma(\varepsilon_{1})-F.
\label{eq:Conditions_load}
\end{equation}
The results of these simulations with $F=5$ and the other  parameters as in the first set of tests  are summarized in Fig.~\ref{fig:Numerics_2}.  We again observe the formation of the transformation front AB moving with velocity   $V=1.42$ and we have checked that this value does not depend on the magnitude of the applied   force. The  inset on the right in Fig.~\ref{fig:Numerics_2} shows an excellent agreement  between numerical and analytical  results. As in our other tests  the formation of this nonlinear front is 'compensated' by the formation of a linear  wave which propagates with  $V=1$ and $\varepsilon_C=F-\sigma_0$  and whose  structure  is  analytically characterized in appendix~\ref{sec:Sonic wave}.   

\section{Quasi-continuum model}
\label{sec:Quasi-continuum model}
 It is of interest to check  to what extent  the simplest  mesoscopic quasi-continuum model,  providing only a  long wavelength approximation of the microscopic  discrete solution, and capturing only  some of its dispersive properties, is compatible with the obtained lattice scale solution. 
 
If we  use the first nontrivial term in the  Fourier space polynomial expansion of the linear operator involved in the formulation of the discrete theory \cite{truskinovsky2006quasicontinuum,slepyan2012models}   we obtain  a  continuum  model with the elastic energy 'corrected'  by the strain gradient term. The kinetic energy remains the same as in conventional continuum theory and the total energy can be written in the form 
 \begin{equation}
 \label{eq:Quasi-contiuum_displ1}
\mathcal{U}=\int_{-\infty}^{\infty}\left[\frac{\rho v^2}{2}+\phi(\varepsilon)-\frac{Ea^2}{12}\left(\frac{\partial \varepsilon}{\partial x}\right)^2\right]\,dx
\end{equation}
where the factor $a^2/12$ reflects the fact that the discrete model contained nearest neighbor interactions only~\cite{mindlin1968theories}. 
Note that the strain gradient term appears with the negative sign which makes the  model  \eqref{eq:Quasi-contiuum_displ1} unstable with respect to perturbations with sufficiently small wave lengths. However, such unstable wave lengths will be absent in the  solutions obtained below which suggests that the predictions of this model can still be trusted.

The  quasi-continuum analog of the  equation \eqref{eq:ShockProblem2}  takes the form 
\begin{equation}
\rho\frac{\partial^2 u(x,t)}{\partial t^2}=\frac{\partial \sigma(\varepsilon)}{\partial x}+\frac{E a^2}{12}\frac{\partial^4 u(x,t)}{\partial x^4},
\label{eq:Quasi-contiuum_displ}
\end{equation}
where the dependence $\sigma=\sigma(\varepsilon)$ is given by \eqref{eq:Stress_Energy}. 
\begin{figure}[h!]
\center{\includegraphics[scale=0.3 ]{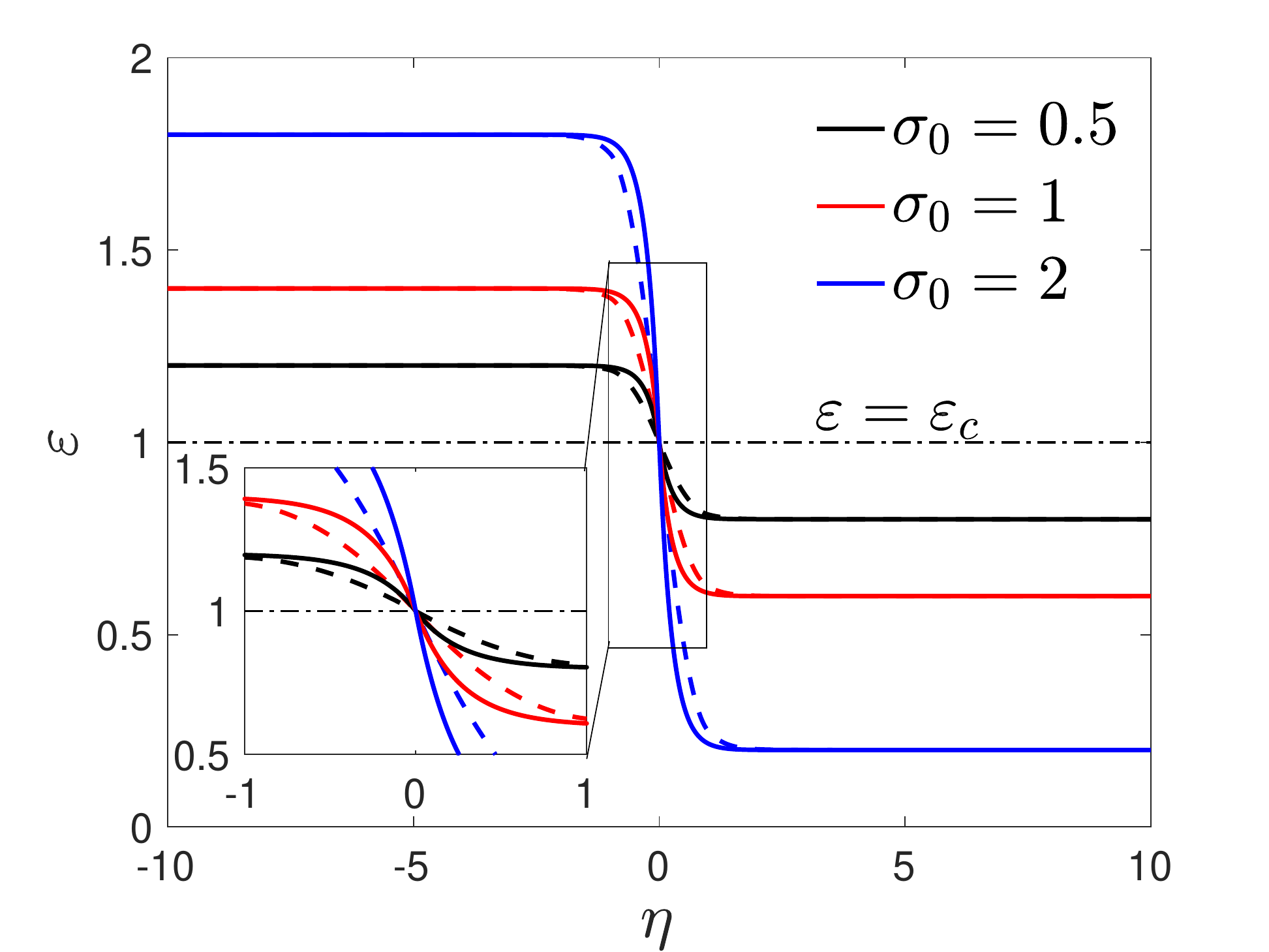}}
\caption[ ]{Quasi-continuum profile of the transformation front  at $V=1.5$ and different $\sigma_0$ (solid lines). The dashed lines show  corresponding solutions of the   discrete problem.}
\label{fig:Approx}
\end{figure}
If we  nondimensionalize  \eqref{eq:Quasi-contiuum_displ}  and  limit our attention to traveling waves we obtain
\begin{equation}
V^2\frac{d^2\varepsilon(\eta)}{d\eta^2}=\frac{d^2\sigma(\eta)}{d\eta^2}+\frac{1}{12}\frac{d^4\varepsilon(\eta)}{d\eta^4}.
\end{equation}
The solution of this equation  can be again obtained using the Fourier transform 
\begin{equation}
\label{fig:Approx1}
\varepsilon(\eta)=\varepsilon_++\frac{\sigma_0}{2\pi i}\int_{-\infty}^{\infty}\frac{12}{(p+iz)(p-iz)}\frac{e^{-ip\eta}}{p}\,dp
\end{equation}
where $z=\sqrt{12(V^2-1)}$. To compute the integral in \eqref{fig:Approx1} we can  again apply the  residue theorem  remembering that the point $p=0$ should be   passed from below.  In this case we have to deal with  much simpler dispersion  relation
\begin{equation}
\omega^2(p)=p^2-p^4/12. 
\label{fig:quasicont}
\end{equation}
As a result, the  only three singularities of the denominator in \eqref{fig:Approx1} are at   $p=0$ and $p=\pm iz$, see  Fig.~\ref{fig:Dispersion_relation}. The solution can be then written explicitly
\begin{equation}
\label{fig:quasicont1}
\varepsilon(\eta)=\begin{cases}
\varepsilon_++\dfrac{\sigma_0}{2}\dfrac{1}{V^2-1}e^{-z\eta},\quad \eta>0\\
\varepsilon_++\dfrac{\sigma_0}{V^2-1}-\dfrac{\sigma_0}{2}\dfrac{1}{V^2-1}e^{z\eta},\quad \eta<0,
\end{cases}
\end{equation}
One can check that the average strains are related again through the RH condition:
$
\varepsilon_-=\varepsilon_++ \sigma_0/(V^2-1).
$
To find  the velocity distribution  we can use the equation $v(\eta)=-V\varepsilon(\eta)$ which gives: 
\begin{equation}
v(\eta)=\begin{cases}
-V\varepsilon_+-\dfrac{\sigma_0}{2}\dfrac{V}{V^2-1}e^{-z\eta},\quad \eta>0,\\
-V\varepsilon_-+\dfrac{\sigma_0}{2}\dfrac{V}{V^2-1}e^{z\eta},\quad \eta<0.
\end{cases}
\end{equation}
The second RH condition is now satisfied automatically. 

If we now require the solution \eqref{fig:quasicont1} to meet the switching  condition   at $\eta=0$ we obtain the same   (trivial) kinetic relation as in the discrete case,  $G(V)= 0$. The absence of dissipation is again due to the supersonic nature of our kinks: this leads to  the absence of   real roots of the quasi-continuum characteristic relation which could describe linear waves carrying energy away from the moving kink.

Examples of the quasi-continuum strain profiles  at $V=1.5$ are shown in Fig.~\ref{fig:Approx} by solid lines. In the same figure we also  show the associated  solutions of the discrete problem (dashed lines). The inset illustrates  the mismatch which is visible only in the immediate vicinity of the origin. 

We can conclude that the  quasi-continuum description of the supersonic active kinks based on \eqref{eq:Quasi-contiuum_displ} is fully adequate and is missing only the fine details of the structure of the core region of the moving front. We recall that to capture these fine details  we needed to account for the  infinite number of the complex roots of the discrete  characteristic equation. Also recall that  the corresponding fully adequate solution was obtained in the form of infinite series. Instead, the quasi-continuum solution, which relies only on a small number of roots,  captures the main features of the discrete model  while remaining  not only explicit but also  extremely simple.  Similarly remarkable efficiency of the  simple quasi-continuum approximations in the description of supersonic solitons in  discrete chains was  shown in \cite{truskinovsky2014solitary}.

\section{Active solitary waves}
\label{sec:Active solitary waves}
To further illustrate the discrete model  we now construct analytically   approximate  solitary wave solutions describing autonomously propagating activity bands. We then demonstrate  numerically that such solutions exhibit  some of the  collision features  characteristic of actual solitons.  

Observe that  in view of its non-dissipative nature, the frontal kink, transforming a passive state  into an active state  and sustained by an internal source  with intensity $\Delta>0$,  can be in principle followed by a symmetry related,   equally non-dissipative rear anti-kink which requires for its propagation the  energy removal performed  by an internal  sink with intensity $-\Delta<0$. We assume that such sources and sinks are present: the discussion of their   micro-realization  goes beyond the scope in this paper.
 
More specifically, in §~\ref{sec:Discrete model} we obtained a traveling wave solution describing  the transition from the state A to the state B which advances at velocity $V$. We can now construct  a traveling wave solution describing transition from B to A which moves with the same velocity. The constants in the boundary conditions \eqref{eq:BCs} should be simply swapped and we obtain the following relations for the strain field 
\begin{equation}
\begin{gathered}
\varepsilon(\eta)=
\begin{cases}
\varepsilon_--\sum\limits_{p_j\in Z^-}\dfrac{\sigma_0\omega^2(p_j)}{p_jL'(p_j)}e^{-ip_j\eta},\quad \eta>0,\\
\varepsilon_++\sum\limits_{p_j\in Z^+}\dfrac{\sigma_0\omega^2(p_j)}{p_jL'(p_j)}e^{-ip_j\eta},\quad \eta<0,
\end{cases}
\end{gathered}
\label{eq:Solution_inv}
\end{equation}
the velocity field
\begin{equation}
v(\eta)=
\begin{cases}
-V\varepsilon_-+\frac{V}{2}\sum\limits_{p_j\in Z^-}\dfrac{\sigma_0\omega^2(p_j)}{\sin{(p_j/2)}L'(p_j)}e^{-ip_j(\eta-1/2)},\quad \eta>1/2,\\
-V\varepsilon_+-\frac{V}{2}\sum\limits_{p_j\in Z^+}\dfrac{\sigma_0\omega^2(p_j)}{\sin{(p_j/2)}L'(p_j)}e^{-ip_j(\eta-1/2)},\quad \eta<1/2,
\end{cases}
\label{eq:Solution_velocity_inv}
\end{equation}
and the displacement field
\begin{equation}
u(\eta)=
\begin{cases}
\varepsilon_-\eta+\dfrac{\sigma_0}{2(V^2-1)} -\dfrac{i}{2}\sum\limits_{p_j\in Z^-}\dfrac{\sigma_0\omega^2(p_j)}{p_j\sin{(p_j/2)}L'(p_j)}e^{-ip_j(\eta-1/2)},\quad \eta>1/2,\\
\varepsilon_+\eta+\dfrac{i}{2}\sum\limits_{p_j\in Z^+}\dfrac{\sigma_0\omega^2(p_j)}{p_j\sin{(p_j/2)}L'(p_j)}e^{-ip_j(\eta-1/2)},\quad \eta<1/2.
\end{cases}
\label{eq:Solution_displ_inv}
\end{equation} 
The equation of the energy balance \eqref{eq:EntropyCondition}  takes the form:
\begin{equation}
\sigma v \bigr\rvert^{+\infty}_{-\infty} - V \Delta - \frac{d}{dt}\int_{-\infty}^{\infty}\left[\frac{v^2}{2}+\phi(\varepsilon)\right]=GV. 
\label{eq:EntropyCondition_inv}
\end{equation}
Note that  the active energy source   transformed here into the energy sink; the kinetic equation here  is again  $G(V)= 0$. 

\begin{figure}[h!]
\center{\includegraphics[scale=0.3]{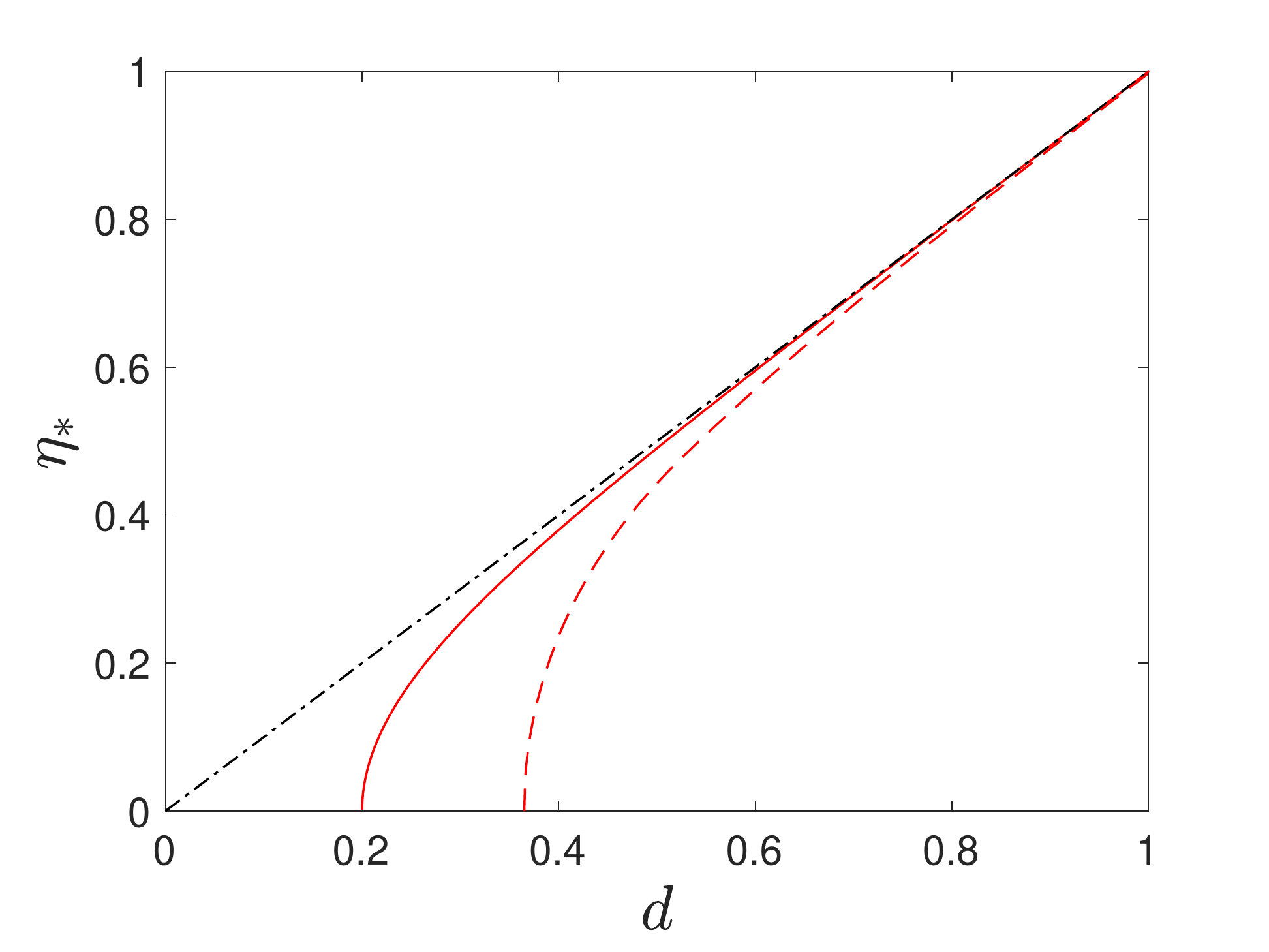}}
\caption[ ]{Solutions of the equation $\varepsilon_{s}(\eta_*)=\varepsilon_c$ for the discrete (red dashed line) and quasi-continuum (red solid line) models with $V=\sqrt{2},\,\varepsilon_c=1,\,\sigma_0=2$. The straight black dash-dotted line is  $\eta_*=d$.}
\label{fig:Eta_star}
\end{figure}

If we now assume that the transition AB takes place  at the value of the TW coordinate  $\eta=d$ and that  the reverse transition BA takes place at $\eta=-d$,  and that in \eqref{eq:equations_steady} the  right-hand side has been  changed to $\sigma(\eta)=\varepsilon(\eta)+\sigma_0\left[H(d-\eta)-H(d+\eta)\right]$,  the approximate solitary wave  solution can be written as a simple superposition of the  corresponding kink and anti-kink solutions:
\begin{equation}
\begin{gathered}
\varepsilon_{s}(\eta)=\varepsilon_++\varepsilon(\eta-d)-\varepsilon(\eta+d),\\
v_{s}(\eta)=v_++v(\eta-d)-v(\eta+d),\\
u_{s}(\eta)=\varepsilon_+\eta+u(\eta-d)-u(\eta+d).
\end{gathered}
\label{eq:Solution_solitons}
\end{equation}
Here  the functions $\varepsilon(\eta)$,  $v(\eta)$ and $u(\eta)$  are given by \eqref{eq:Solution_series}, \eqref{eq:Solution_velocity} and \eqref{eq:Solution_displ};  the constant    $\varepsilon_+$ is given  again by \eqref{eq:Epsilon_pm}. 
\begin{figure}[h!]
\center{\includegraphics[scale=0.3]{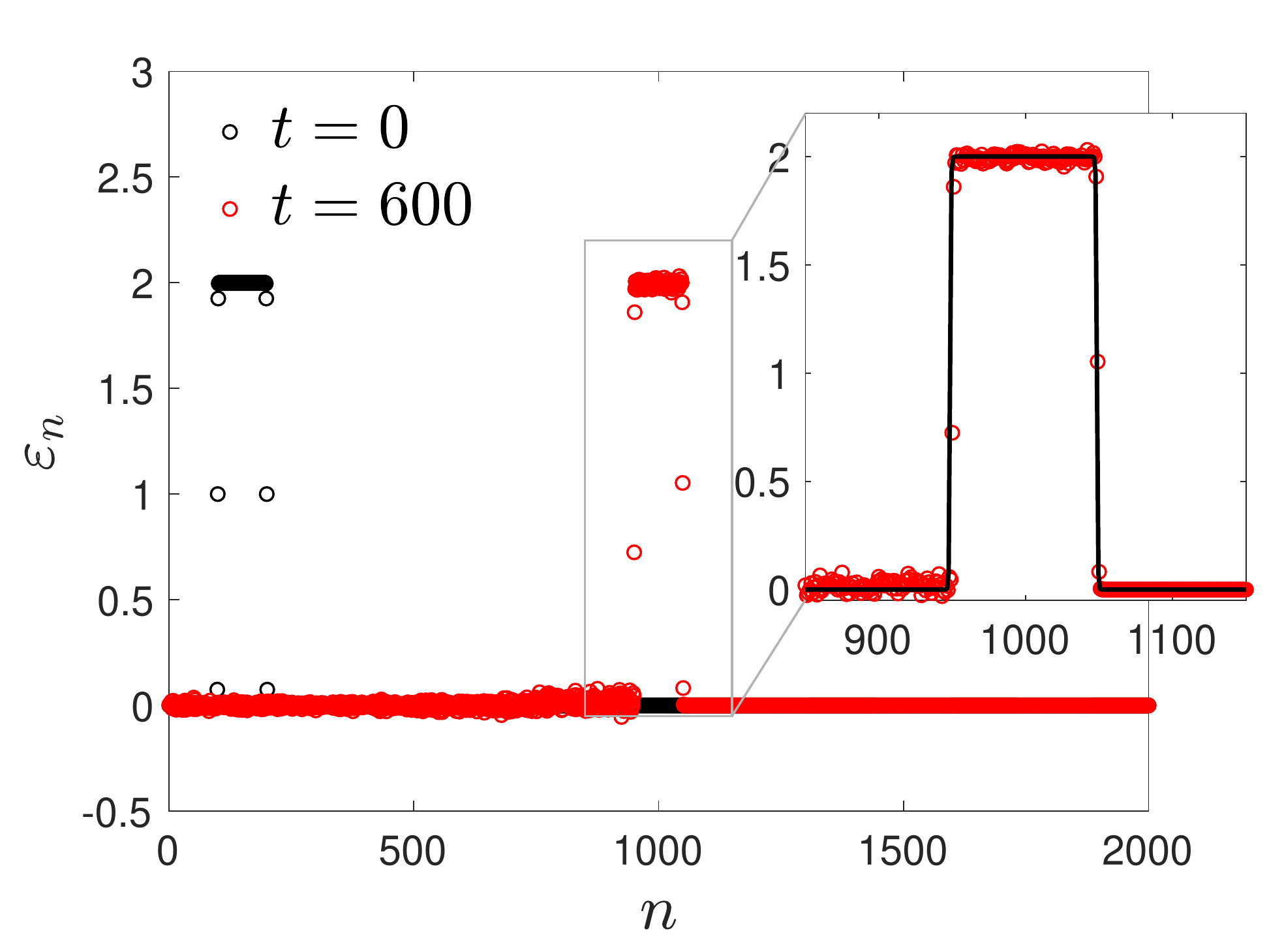}}
\caption[ ]{Snapshots at $t=0$ and $t=600$  of the soliton-type solution  with the half-width $d=50$ propagating at the speed $V=\sqrt{2}$. The inset compares the  initial and the current configurations.}
\label{fig:Strain_Soliton}
\end{figure}
 
Suppose next that we fix $d$ and solve the algebraic equation $\varepsilon_{s}(\eta_*)=\varepsilon_c$ obtaining the  relation $\eta_*(d)$. The traveling wave  \eqref{eq:Solution_solitons} is an exact  solution  if  there exists a value of $d$ such that  $\eta_*(d)=d$. The function $\eta_*=\eta_*(d)$ computed numerically  is shown in Fig.~\ref{fig:Eta_star}   (red dashed line)  where we also present its quasicontinuum analog (red solid line) which is known analytically 
\begin{equation}
\eta_*= z^{-1} \cosh^{-1}\left((\varepsilon_c-\varepsilon_+)\left[1-\dfrac{V^2-1}{\sigma_0}\right]\exp{(zd)}\right),
\end{equation} 
and where $z=\sqrt{12(V^2-1)}$. While none of these curves crosses the straight line $\eta =d$, they become close at large $d$ which suggests that in this range of parameters  the approximate configuration \eqref{eq:Solution_solitons} will evolve towards the actual solitary wave type solution (if it exists).

\begin{figure}[h!]
\center{\includegraphics[scale=0.25]{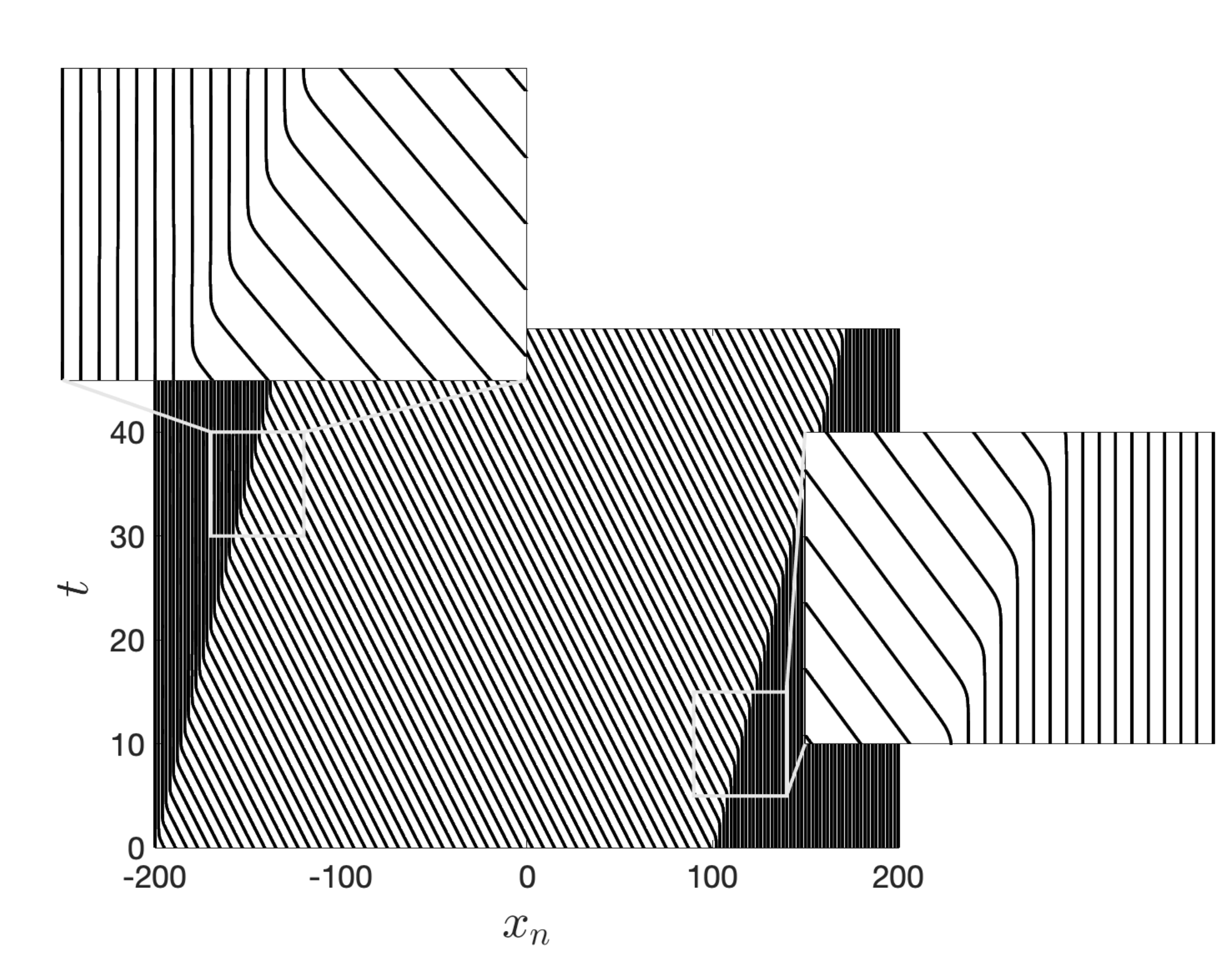}}
\caption[ ]{Particle motion  $x_n(t)=n+u_n(t)$ associated with the soliton-type solution shown in Fig. \ref{fig:Strain_Soliton} for $n=100,102,104,...,200$.}
\label{fig:Position_Soliton}
\end{figure}

To check the actual existence of such solitary waves,  we resolve to numerical simulations. We take   the ansatz \eqref{eq:Solution_solitons} as our initial data,   keep the parameters  $\sigma_0=2,\,\varepsilon_c=1$ as in our  previous numerical simulations and set $V=\sqrt{2}$ which results in $\varepsilon_+=0$. 

 If we choose initial perturbation with small  $d \sim 1$, when the approximation is expected to be poor,  the ensuing dynamics  shows   de-localization of the pulse with  fast decay of its amplitude. 

If, instead, the initial pulse is wide with $d \sim 50$ its evolution  is completely different, see  Fig.~\ref{fig:Strain_Soliton}. Our numerical simulations show a steadily propagating solitary wave with the speed anticipated by our analytical approximation even though we also observe some noise left in the wake. We can conjecture that  even if initially 
localized energy   eventually dissipates into small scale  oscillations, it  would take considerable time.    Note that the resulting motion,  see Fig.~\ref{fig:Position_Soliton},  is supported exclusively by the internal activity which effectively \emph{translocates}  the particles  to the left by $\sim 200$ units.

\begin{figure}[h!]
\begin{center}
a)\hspace{-0.5cm}\begin{minipage}[h]{0.35\linewidth}
\center{\includegraphics[scale=0.25]{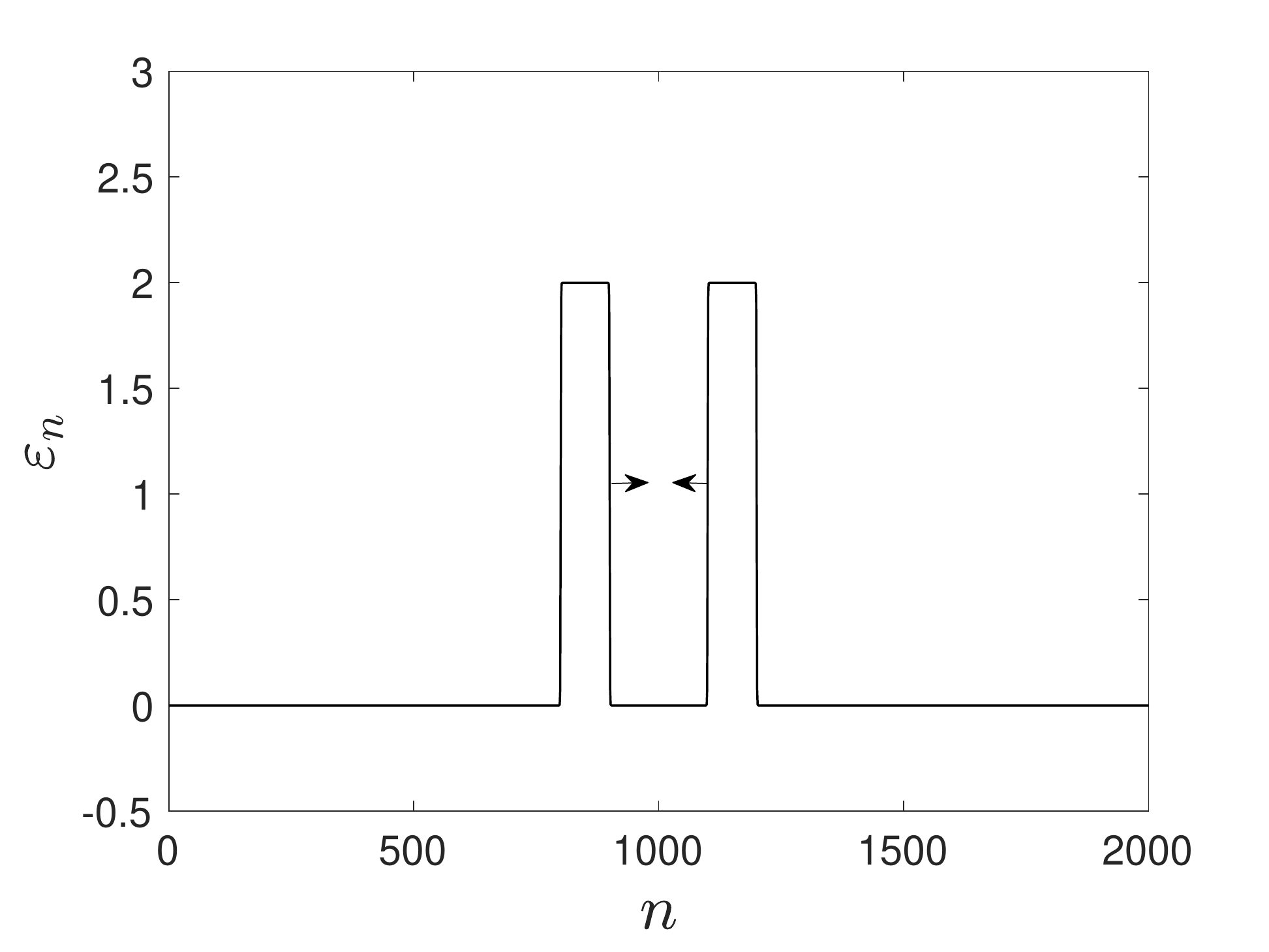}}
\end{minipage}
b)\hspace{-0.5cm}\begin{minipage}[h]{0.35\linewidth}
\center{\includegraphics[scale=0.25]{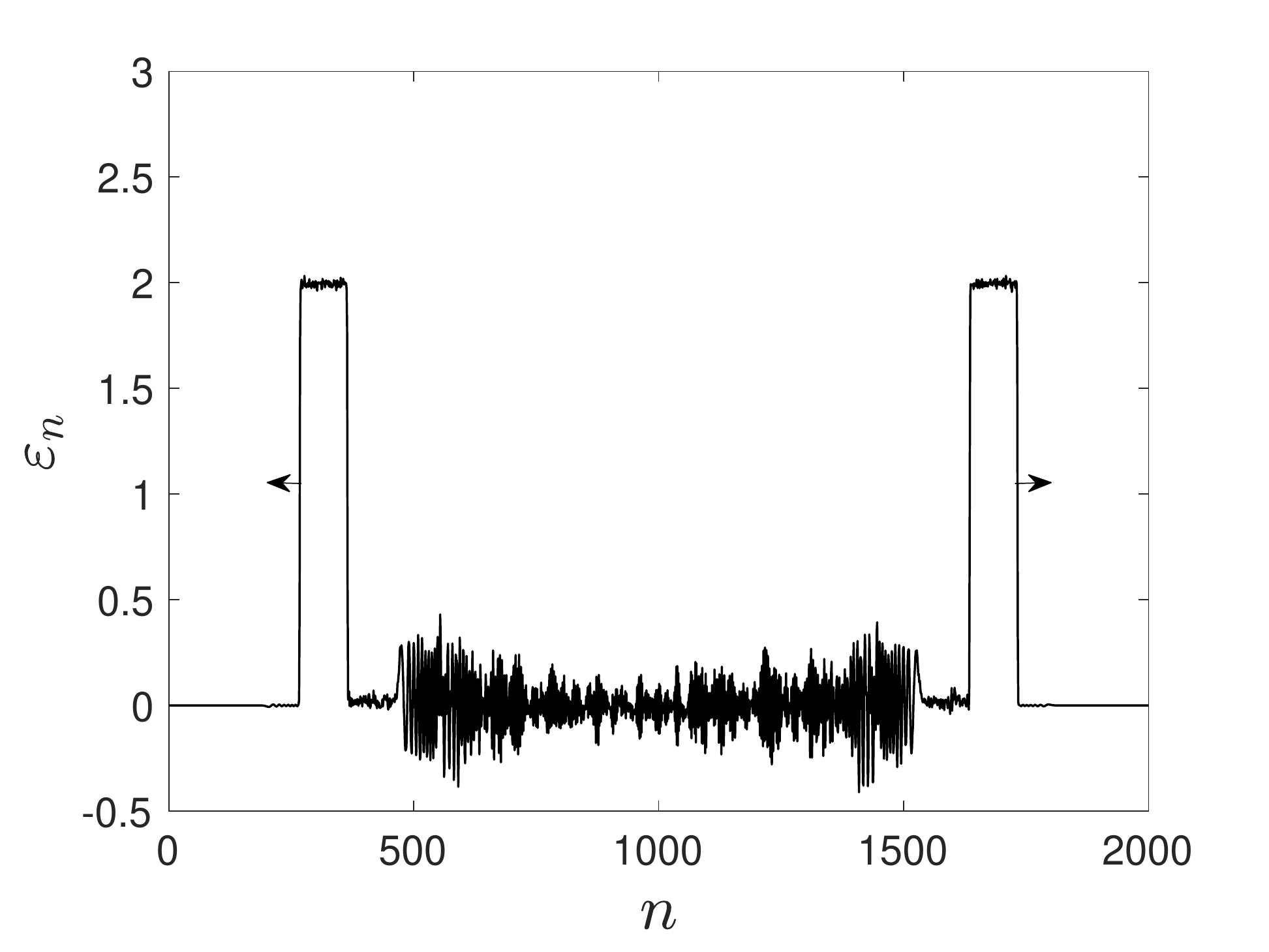}}
\end{minipage}
\end{center}
\caption[ ]{Quasi-elastic collision of  two solitary waves  with  half-width $d=50$ and speed $V=\sqrt{2}$: a) before the collision at $t=0$, b) after the collision at $t=600$. Arrows indicate the propagation directions.}
\label{fig:Collision}
\end{figure}

Finally, to show that the obtained solitary wave type solutions have some properties of actual solitons, we performed   the   collision between  two such waves  with half-width $d=50$ moving against each other with velocities  $V= \pm \sqrt{2}$, see   Fig.~\ref{fig:Collision}.   We observe that despite almost elastic interaction,  a very limited amount of energy is lost  because of the formation of  small scale oscillations. In the Supplementary Material we included the animation showing that  during the collision, taking place between  $t\approx 60$ and  $t\approx 150$, there is a considerable  increase of strain which then subsides to the values  prescribed  by the initial data.

\section{Conclusions}
\label{sec:Conclusions}

We used the simplest model of an active solid to obtain an analytical  description of a propagating front  which transform  a passive phase (no active stress) into  an active phase (active stress is present).  Inertial effects were taken into consideration because the targeted solids are  expected to be anomalously \emph{soft}.

We assumed that the steady advancement  of such  fronts can be self-sustained in the sense that it can be fuelled by an  internal energy reservoir  presenting itself as  an effective source of  \emph{anti-dissipation}.  In various biological settings the implied activity is revealed through the work done, for instance, by the  macroscopically invisible molecular motors.  Our conclusion that the transformation fronts  must be necessarily supersonic  is meaningful because in  biologically relevant   active solids  the acoustic speeds may be arbitrarily small.

As we showed, the supersonic nature of the transformation fronts  eliminates the possibility of the radiative damping which is an important source of dissipation for classical defects in crystalline solids. Such  purely mechanical dissipation can be therefore neglected in the case of propagating activity fronts which emerge as largely anti-dissipative. 

We presented three different descriptions of such transformation fronts: continuum, discrete and quasicontinuum. The classical continuum model is not self-contained because the condition of non-dissipativeness has to be added phenomenologically. Instead, both discrete and quasi-continuum models provide this supplementary condition automatically. 

Since our discrete model basically coincides with the well known FPU system,  we effectively generalized the latter to the case when springs can be both passive and  active,  allowing for the propagation of  active phase transition fronts.  Our ability to  construct quasi-continuum solutions reproducing faithfully  all the  properties as the discrete solution shows that the complexity of the dispersion relations exhibited by the discrete system is not necessary for capturing the observed  behavior which can be already reproduced  using the simplest polynomial approximation of the discrete dispersion relation. 
 Our quasi-continuum approximation coincides  with the 'bad Boussinesque' model which is generically unstable but works in our case since the obtained solution is confined to  a finite ball  in the Fourier space where the implied instability is absent.

From the perspective of the theory of  nonlinear waves in dispersive systems the obtained  kink type solutions are special in the sense that  the discreteness of the structure does not automatically lead to lattice resonances and the associated dissipation. Such nondissipative propagation of lattice ways is sometimes interpreted as lattice transparency. While in subsonic case such regimes  require very particular conditions, in the supersonic case  they  becomes  robust.
 
\subsection*{Funding}{NG acknowledges the support of the French Agence Nationale de la Recherche (ANR) under reference ANR-17-CE08-0047-02. LT  was supported by  the grant  ANR-10-IDEX-0001-02 PSL. }

\subsection*{Acknowledgements}{ We dedicate this work to the 90th birthday of Leonid Slepyan, an amazing scientist and a truly great man. 
 }
 
\appendix
\section{Appendix}
\label{sec:Sonic wave}

Following \cite{slepyan1972nonstationary}, we consider the linear problem (no transition). We start with the solution for the semi-infinite domain which  can be obtained by solving the problem symmetric loading
\begin{equation}
\frac{d^2 u(x,t)}{d t^2}=u(x+1,t)+u(x-1,t)-2u(x,t)-\delta(x-1)-\delta(x),\quad -\infty<x<\infty,
\label{eq:FundamentalProblem}
\end{equation}
and then taking $x\geq1$.  It is implied here  that the initial conditions are  trivial: $u(x,0)=0,du(x,0)/dt=0$. Because of the symmetry $\varepsilon(1,t)=u(1,t)-u(0,t)=0$ and  by inverting  the discrete Fourier transform, we obtain the desired solution
\begin{equation}
\varepsilon(x,t)=\frac{2}{\pi}\int_{0}^{\pi/2}\frac{\cos{p}}{\sin{p}}(1-\cos{(2t\sin{p})})\sin{2px}\,dp.
\label{eq:SolutionGreen}
\end{equation}
To match the transitions BC or DC shown in Fig.~\ref{fig:Numerics_1} and Fig.~\ref{fig:Numerics_2}, we need to multiply \eqref{eq:SolutionGreen} by the appropriate factor and adjust the shift, for instance, to match the transition BC in Fig.~\ref{fig:Numerics_1} we need to use the multiple $(\varepsilon_C-\varepsilon_+)$ and the additive term $\varepsilon_+$.  After these adjustments we  obtain the profiles shown in the insets in Fig.~\ref{fig:Numerics_1} and Fig.~\ref{fig:Numerics_2} (magenta lines).  

To obtain analytical results  we need to study the integral in \eqref{eq:SolutionGreen} in the limit $t\to\infty$, see also \cite[sec.~{I}]{slepyan1972nonstationary}.  We first rewrite it in the form:
\begin{equation}
\begin{gathered}
\varepsilon(x,t)=-\frac{1}{\pi}\int_0^{\pi/2}\frac{\cos{p}}{\sin{p}}\left[\sin{(2(px+t\sin{p}))}+\sin{(2(px-t\sin{p}))}\right]\,dp\\
+\frac{2}{\pi}\int_{0}^{\pi/2}\frac{\cos{p}}{\sin{p}}\sin{2px}\,dp.
\end{gathered}
\label{eq:Solution_integrals}
\end{equation}
The time evolution is contained in the first integral where the phases $px+t\sin{p}$ and $px-t\sin{p}$ describe waves  propagating in the negative and positive $x$ directions, respectively.The last integral is the irrelevant time independent term.

In the limit $t\to\infty$ the major contribution to the integral comes from the end points of the interval and at stationary points of the phases. However, the integrand is zero at $p=\pi/2$ due to the presence of the term $\cos{p}$. Moreover, the stationary points $p_{1,2}$ are such that $\cos{p_{1,2}}=\pm p/t$. This becomes  $\pm \pi/2$ in the limit $t\to\infty$ and, hence, the  contribution of these terms  is  minor. Therefore, it is sufficient  to account for  the contribution to the integral  from the points  close to $p=0$:
\begin{equation}
\varepsilon(x,t)\sim -\frac{1}{\pi}\int_0^{\delta}\frac{\sin{(2(px+t\sin{p}))}+\sin{(2(px-t\sin{p}))}}{p}\,dp,\quad t\to\infty
\end{equation}
where we introduced the small parameter $\delta\ll 1$. If $|x\pm t|\geq t^a,\,a>1/3$, we can make the  change of variables $y=p(x\pm t)$. Then in the limit  $t\to\infty$ the upper integration limit becomes infinity and the integrals converge to $\pi/2\,\text{sign}{(x\pm t)}$.

A different result is obtained  when $|x\pm t|=O(t^{1/3})$. Making the change of variables $z=p t^{1/3}$ in the integral and setting $\delta t^{1/3}\to\infty$ we get:
\begin{equation}
\int_0^{\delta}\frac{\sin{(2(x\pm t)p\mp p^3 t/3)}}{p}=\int_0^{\infty}\frac{\sin{(\xi^{\pm}z\mp z^3/3})}{z}\,dz,\quad \xi^\pm=2\frac{x\pm t}{t^{1/3}}
\end{equation}
\begin{figure}[h!]
\center{\includegraphics[scale=0.25]{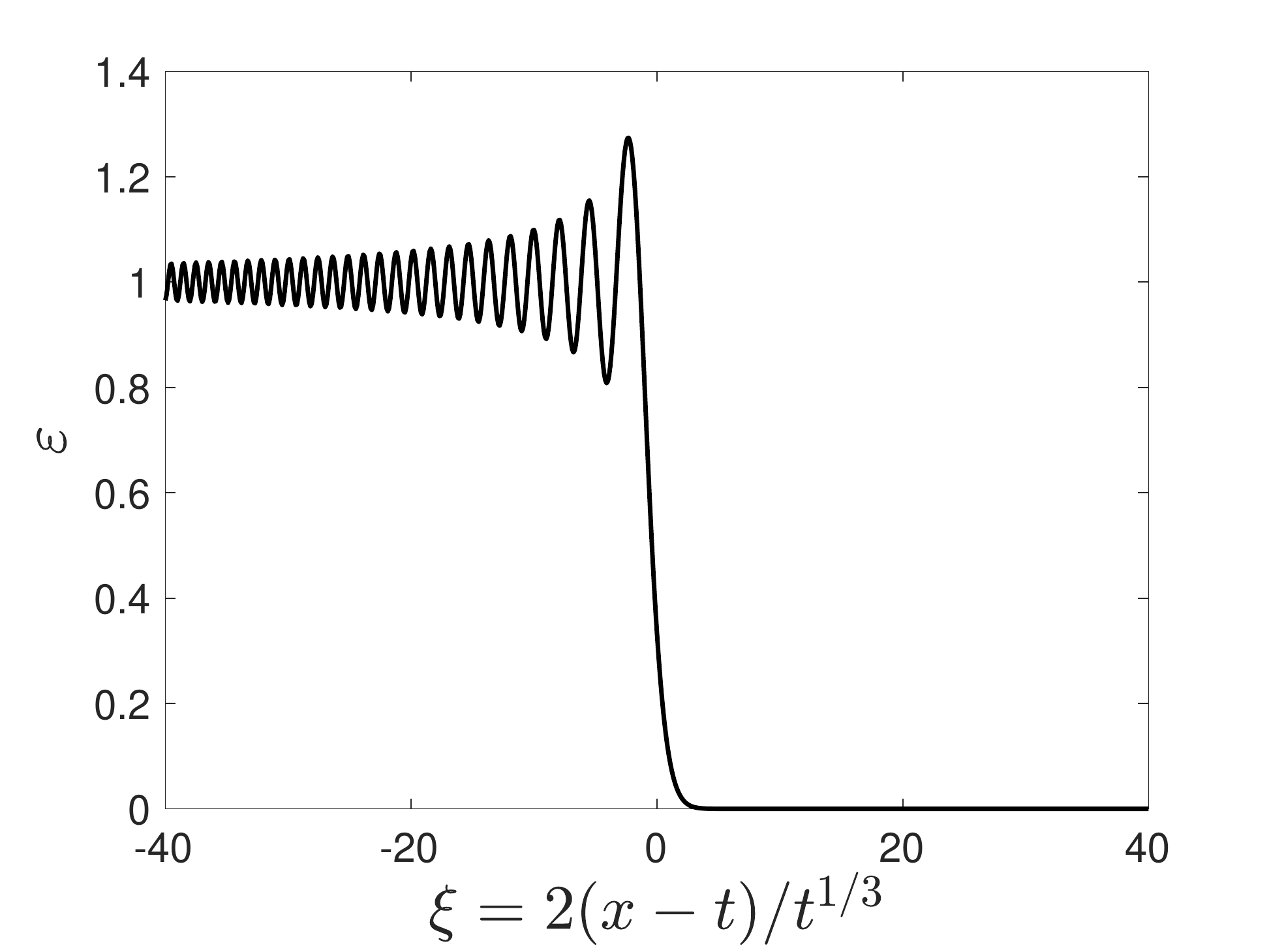}}
\caption[ ]{Asymptotic behaviour \eqref{eq:Sonic_asymp} for the sonic wave when $t\to\infty$.}
\label{fig:Asymptotic_sonic}
\end{figure}
From these expressions we see that the sonic wave propagates with velocity $V=1$ and its front spreading follows the asymptotics  $t^{1/3}$. More precisely, we can write 
\begin{equation}
\varepsilon(x,t)\sim \int_{\xi^-}^{\infty} \text{Ai}(x)dx+\int_{\xi^+}^{\infty} \text{Ai}(-x)dx,\quad t\to\infty,\quad \text{Ai}(x)=\frac{1}{\pi}\int_0^{\infty}\cos{\left(zx+\frac{z^3}{3}\right)}\,dz,
\end{equation}
where $\text{Ai}(x)$ is the Airy function. Finally, by letting $\xi^+\to\infty$, i.e. by switching to the moving frame $\xi^-$, we can make the  second integral vanish and the asymptotic expression for strain takes a simple form:
\begin{equation}
\varepsilon(x,t)\sim \int_{\xi}^{\infty} \text{Ai}(x)dx,\quad t\to\infty,\quad \xi=2\frac{x-t}{t^{1/3}}
\label{eq:Sonic_asymp}
\end{equation}
Function \eqref{eq:Sonic_asymp} is plotted in Fig.~\ref{fig:Asymptotic_sonic}. The obtained profile is the same as the ones emerging in our  numerical simulations, see Fig.~\ref{fig:Numerics_1} and Fig.~\ref{fig:Numerics_2}. The exact matching can be achieved  by  shifting the origin and adding  the  appropriate factor.

\clearpage
\bibliography{Bibliography}{}

\begin{thebibliography}{10}

\bibitem{marchetti2013hydrodynamics}
MC~Marchetti, J-F Joanny, S~Ramaswamy, TB~Liverpool, J~Prost, M~Rao, and
  RA~Simha.
\newblock Hydrodynamics of soft active matter.
\newblock {\em Reviews of Modern Physics}, 85(3):1143, 2013.

\bibitem{de2015introduction}
G~De~Magistris and D~Marenduzzo.
\newblock An introduction to the physics of active matter.
\newblock {\em Physica A: Statistical Mechanics and its Applications},
  418:65--77, 2015.

\bibitem{fodor2018statistical}
{\'E}~Fodor and MC~Marchetti.
\newblock The statistical physics of active matter: From self-catalytic
  colloids to living cells.
\newblock {\em Physica A: Statistical Mechanics and its Applications},
  504:106--120, 2018.

\bibitem{berthier2019glassy}
L~Berthier, E~Flenner, and G~Szamel.
\newblock Glassy dynamics in dense systems of active particles.
\newblock {\em The Journal of chemical physics}, 150(20):200901, 2019.

\bibitem{solon2015pattern}
AP~Solon, J-B Caussin, D~Bartolo, H~Chat{\'e}, and J~Tailleur.
\newblock Pattern formation in flocking models: A hydrodynamic description.
\newblock {\em Physical Review E}, 92(6):062111, 2015.

\bibitem{ngamsaad2018propagating}
W~Ngamsaad and S~Suantai.
\newblock Propagating wave in the flock of self-propelled particles.
\newblock {\em Physical Review E}, 98(6):062618, 2018.

\bibitem{prost2015active}
J~Prost, F~J{\"u}licher, and J-F Joanny.
\newblock Active gel physics.
\newblock {\em Nature Physics}, 11(2):111, 2015.

\bibitem{julicher2018hydrodynamic}
F~J{\"u}licher, SW~Grill, and G~Salbreux.
\newblock Hydrodynamic theory of active matter.
\newblock {\em Reports on Progress in Physics}, 81(7):076601, 2018.

\bibitem{hawkins2014stress}
RJ~Hawkins and TB~Liverpool.
\newblock Stress reorganization and response in active solids.
\newblock {\em Physical review letters}, 113(2):028102, 2014.

\bibitem{maitra2018oriented}
A~Maitra and S~Ramaswamy.
\newblock Oriented active solids.
\newblock {\em arXiv preprint arXiv:1812.01374}, 2018.

\bibitem{moshe2018geometric}
M~Moshe, MJ~Bowick, and MC~Marchetti.
\newblock Geometric frustration and solid-solid transitions in model 2{D}
  tissue.
\newblock {\em Physical review letters}, 120(26):268105, 2018.

\bibitem{scheibner2019odd}
C~Scheibner, A~Souslov, D~Banerjee, P~Surowka, W~Irvine, and V~Vitelli.
\newblock Odd elasticity in active metamaterials.
\newblock {\em arXiv preprint arXiv:1902.07760}, 2019.

\bibitem{finlayson1969convective}
BA~Finlayson and LE~Scriven.
\newblock Convective instability by active stress.
\newblock {\em Proceedings of the Royal Society of London. A. Mathematical and
  Physical Sciences}, 310(1501):183--219, 1969.

\bibitem{cohen1991paradoxical}
JE~Cohen and P~Horowitz.
\newblock Paradoxical behaviour of mechanical and electrical networks.
\newblock {\em Nature}, 352(6337):699, 1991.

\bibitem{braess1968paradoxon}
D~Braess.
\newblock {\"U}ber ein paradoxon aus der verkehrsplanung.
\newblock {\em Unternehmensforschung}, 12(1):258--268, 1968.

\bibitem{nicolaou2012mechanical}
ZG~Nicolaou and AE~Motter.
\newblock Mechanical metamaterials with negative compressibility transitions.
\newblock {\em Nature materials}, 11(7):608, 2012.

\bibitem{ayzenberg2014brittle}
M~Ayzenberg-Stepanenko, G~Mishuris, and L~Slepyan.
\newblock Brittle fracture in a periodic structure with internal potential
  energy. spontaneous crack propagation.
\newblock {\em Proceedings of the Royal Society A: Mathematical, Physical and
  Engineering Sciences}, 470(2167):20140121, 2014.

\bibitem{gallavotti2007fermi}
G~Gallavotti.
\newblock {\em The {F}ermi-{P}asta-{U}lam problem: a status report}, volume
  728.
\newblock Springer, 2007.

\bibitem{truskinovsky2005kinetics}
L~Truskinovsky and A~Vainchtein.
\newblock Kinetics of martensitic phase transitions: lattice model.
\newblock {\em SIAM Journal on Applied Mathematics}, 66(2):533--553, 2005.

\bibitem{slepyan2005transition}
L~Slepyan, A~Cherkaev, and E~Cherkaev.
\newblock Transition waves in bistable structures. {II}. {A}nalytical solution:
  wave speed and energy dissipation.
\newblock {\em Journal of the Mechanics and Physics of Solids}, 53(2):407--436,
  2005.

\bibitem{truskinovsky2006quasicontinuum}
L~Truskinovsky and A~Vainchtein.
\newblock Quasicontinuum models of dynamic phase transitions.
\newblock {\em Continuum Mechanics and Thermodynamics}, 18(1-2):1--21, 2006.

\bibitem{slepyan1988impact}
LI~Slepyan and LV~Troyankina.
\newblock Impact waves in a nonlinear chain.
\newblock {\em Plasticity and Fracture of Solids}, pages 175--186, 1988.
\newblock (In Russian).

\bibitem{slepyan2012models}
LI~Slepyan.
\newblock {\em Models and phenomena in fracture mechanics}.
\newblock Springer Science \& Business Media, 2012.

\bibitem{truskinovskii1987dynamics}
L~Truskinovskii.
\newblock Dynamics of non-equilibrium phase boundaries in a heat conducting
  non-linearly elastic medium.
\newblock {\em Journal of Applied Mathematics and Mechanics}, 51(6):777--784,
  1987.

\bibitem{truskinovsky1993kinks}
L~Truskinovsky.
\newblock Kinks versus shocks.
\newblock In {\em Shock induced transitions and phase structures in general
  media}, pages 185--229. Springer, 1993.

\bibitem{mindlin1968theories}
RD~Mindlin.
\newblock Theories of elastic continua and crystal lattice theories.
\newblock In {\em Mechanics of Generalized Continua}, pages 312--320. Springer,
  1968.

\bibitem{truskinovsky2014solitary}
L~Truskinovsky and A~Vainchtein.
\newblock Solitary waves in a nonintegrable {F}ermi-{P}asta-{U}lam chain.
\newblock {\em Physical Review E}, 90(4):042903, 2014.

\bibitem{slepyan1972nonstationary}
LI~Slepyan.
\newblock Nonstationary elastic waves.
\newblock {\em Sudostroenie, Leningrad}, page 376, 1972.
\newblock (In Russian).

\end{thebibliography}
\bibliographystyle{unsrt}
\end{document}